\documentclass[12pt]{article}
\usepackage{titlesec}
\usepackage{mathrsfs}

\usepackage{graphicx}
\usepackage{dcolumn}
\usepackage{bm}

\usepackage[utf8]{inputenc}
\usepackage[T1]{fontenc}

\usepackage{epsfig}
\usepackage{amsfonts}
\usepackage{amssymb,latexsym}

\usepackage{amsmath}
\usepackage{float}

\usepackage{subcaption}

\usepackage{cleveref}

\usepackage[utf8]{inputenc}
\usepackage[english]{babel}
\usepackage{xcolor}

\newcommand{\angstrom}{\mbox{\normalfont\AA}}

\newcommand{\fa}{\mathfrak{a}}
\newcommand{\fb}{\mathfrak{b}}
\newcommand{\fc}{\mathfrak{c}}

\newcommand{\fK}{\mathfrak{K}}

\newcommand{\bn}{{\mathbf{n}}}

\newcommand{\cH}{\mathcal{H}}
\newcommand{\cE}{\mathcal{E}}

\newcommand{\cG}{\mathcal{G}}

\newcommand{\cO}{\mathcal{O}}
\newcommand{\cP}{\mathcal{P}}

\newcommand{\cT}{\mathcal{T}}

\newcommand{\cX}{\mathcal{X}}

\newcommand{\be}{\begin{equation}}
\newcommand{\ee}{\end{equation}}
\newcommand{\bea}{\begin{eqnarray}}
\newcommand{\eea}{\end{eqnarray}}
\newcommand{\nn}{\nonumber}

\newcommand{\ed}{\end{document}}

\newcommand{\bi}{\begin{itemize}}
\newcommand{\ei}{\end{itemize}}

\newcommand{\bce}{\begin{center}}
\newcommand{\ece}{\end{center}}

\newcommand{\sE}{\mathscr{E}}

\newcommand{\RE}{{\rm Re}}
\newcommand{\IM}{{\rm Im}}

\newcommand{\bvtheta}{\boldsymbol{\vartheta}}

\begin{document}

\title{Spectral singularities and tunable slab lasers\\ with 2D material coating}

\author{Hamed Ghaemi-Dizicheh$^{1,3}$\thanks{h.ghaemidizicheh@lancaster.ac.uk}\,, Ali~Mostafazadeh$^{1,2,} $\thanks{Email Address: amostafazadeh@ku.edu.tr}\,, and Mustafa Sar{\i}saman$^4$\thanks{mustafa.sarisaman@istanbul.edu.tr}\\[6pt]
    Departments of Physics$^1$ and Mathematics$^2$, Ko\c{c} University,\\ 34450 Sar{\i}yer,
    Istanbul, Turkey\\[6pt]
    Department of Physics, Lancaster University,\\ Lancaster LA1 4YB, United Kingdom$^3$\\[6pt]
    Department of Physics, Istanbul University,\\ 34134 Vezneciler,
    Istanbul, Turkey$^4$}

\date{ }
\maketitle

\begin{abstract}

We investigate linear and nonlinear spectral singularities in the transverse electric and transverse magnetic modes of a slab laser consisting of an active planar slab sandwiched between a pair of Graphene or Weyl semimetal thin sheets. The requirement of the presence of linear spectral singularities gives the laser threshold condition while the existence of nonlinear spectral singularities due to an induced weak Kerr nonlinearity allows for computing the laser output intensity in the vicinity of the threshold. The presence of the Graphene and Weyl semimetal sheets introduces additional physical parameters that we can use to tune the output intensity of the laser. We provide a comprehensive study of this phenomenon and report peculiarities of lasing in the TM modes of the slab with Weyl semimetal coatings. In particular, we reveal the existence of a critical angle such that no lasing seems possible for TM modes of the slab with smaller emission angle. Our results suggest that for TM modes with emission angle slightly exceeding the critical angle, the laser output intensity becomes highly sensitive to the physical parameters of the coating.

\medskip
\end{abstract}

\maketitle

\section{Introduction}

Spectral singularities were introduced by mathematicians in the 1950's and studied by them for over seven decades \cite{naimark1960investigation,kemp1958singular,schwartz1960some}. They appear as the points of the continuous spectrum of the Schr\"odinger operator, $\partial_x^2+v(x)$, for a complex potential $v(x)$ that are responsible for certain peculiar mathematical properties of this operator \cite{guseinov}. A more recent examination of the physical meaning of spectral singularities has led to their identification with the (real) scattering energies at which the transmission and reflection amplitudes of the potential diverge \cite{prl-2009}. This has led to a growing interest in the study of their physical aspects 
\cite{longhi2009spectral,longhi2010pt,pra-2011a,pla-2011,prsa-2012,correa2012spectral,pra-2013b,pra-2013d,chaos2013resonant,garcia2014time,ramezani2014unidirectional,li2014complete,pra-2015,wang2016wave,hang2016tunable,kalozoumis2017emitter,pendharker2017pt,Zhang-2017,Ahmed-2018,Jin-2018,Konotop-2018,Midya-2018,Zezyulin-2018,Muller-2018,Li-2019,Konotop-2019a,Konotop-2019b,ap-2019b}.

Optics provides a fertile research field where spectral singularities can be realized. For an optical medium involving regions of gain, they correspond to configurations where the medium begins emiting purely outgoing coherent waves, i.e., it acts as a laser. In other words, the mathematical requirement that the optical potential describing the medium possesses a spectral singularity coincides with the laser threshold condition \cite{pra-2011a}.

Ref.~\cite{prl-2013} extends the notion of spectral singularity for a large class of nonlinear wave equations, and Ref.~\cite{pra-2013} employs the resulting nonlinear spectral singularities to explore the effects of a weak Kerr nonlinearity on the behavior of a simple slab laser. The main outcome of this study is the following expression for the laser output intensity.
    \bea
    I=\dfrac{g-g_{0}}{\varsigma \, g_{0}}~\hat I,
    \label{laser intensity}
    \eea
where $g$ is the slab's gain coefficient, $g_{0}$ is its threshold value, $\varsigma$ is the Kerr coefficient, and $\hat I$ is a real and positive coefficient that depends on the geometry and physical parameters of the system. According to (\ref{laser intensity}), lasing can occur whenever $g$ exceeds $g_{0}$, and in this case, the intensity of the emitted laser light is proportional to $g-g_{0}$. Both of these are common knowledge in laser physics, but here they follow from the purely mathematical condition of the existence of a nonlinear spectral singularity due to an induced Kerr nonlinearity.

Refs.~\cite{pra-2015} and \cite{jo-2017} respectively study the linear and nonlinear spectral singularities in the oblique transverse electric (TE) and  transverse magnetic (TM) modes of a homogeneous slab of active material. This leads to explicit formulas for the threshold gain and laser output intensity, and predicts that lasing in the TM modes with emission angles larger than Brewster's angle is forbidden \cite{jo-2017}.

In the present article, we explore the effects of coating the faces of the slab laser considered in Refs.~\cite{pra-2015,jo-2017} by the two-dimensional (2D) materials having a scalar conductivity. This is in part motivated by the recent work on the $\cP\cT$-symmetric coherent perfect absorbers that are obtained by coating a $\cP\cT$-symmetric bilayer with such materials \cite{sarisaman2018pt,sarisaman2019broadband}. See also \cite{mustafa-JAP}. Our main objective is to investigate prospects of using the physical parameters determining the conductivity of the 2D boundary material as tuning parameters for the slab laser.

Consider an infinite homogeneous and isotropic planar Kerr slab of thickness $L$ with one or both faces coated by an extremely thin layer of a 2D material. Suppose that the slab is exposed to external time-harmonic electromagnetic waves, $ e^{-i\omega t}\vec{E}(\vec{r})$ and $ e^{-i\omega t}\vec{H}(\vec{r})$, as depicted in Fig.~\ref{fig1},
    \begin{figure*}
    \begin{center}
    \includegraphics[scale=0.9]{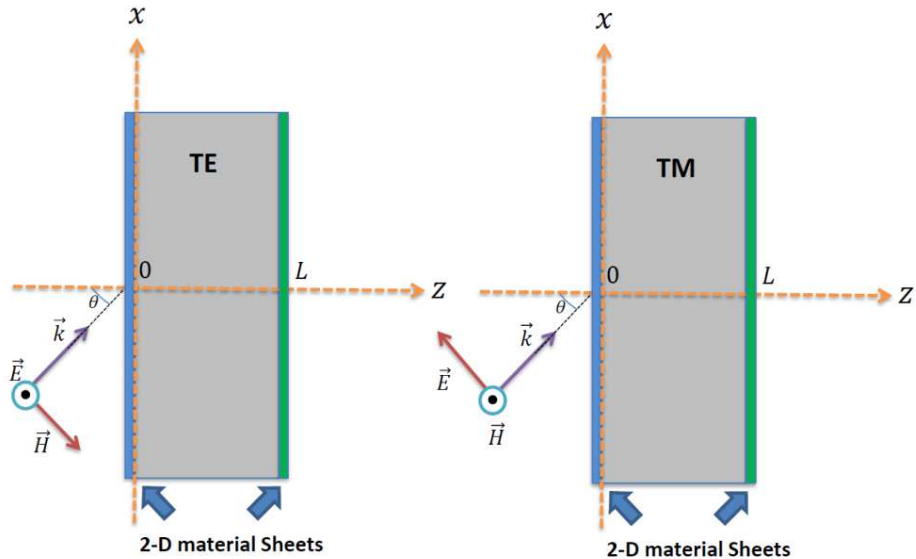}
    \caption{(Color online) Diagrams representing the scattering of the TE (on the left) and $ TM $ (on the right) waves by a planar slab of thickness $L$ which is placed between two-dimensional materials.}
    \end{center}
    \end{figure*} \label{fig1}
and adopt a coordinate system in which the wavevector takes the form $ \vec{k}=k_{x}\hat{e}_{x}+k_{z}\hat{e}_{z} $, where $k_{x}:=k\sin\theta $, $k_{z}:=k\cos\theta $, $ k=\omega/c $ is the wavenumber, $\theta$ is the incidence angle, $\omega$ is the angular frequency, $c$ is the speed of light in vacuum, and $ \hat{e}_{u} $ is the unit vector pointing along the positive $u$-axis with $u=x,y,z$. We wish to explore the effect of the 2D materials coating the faces of the slab on the threshold gain $g_0$ and the laser output intensity $I$ slightly above the threshold. For this purpose we choose 2D materials with a scalar conductivity such as Graphene and Weyl semimetal (WSM) \cite{falkovsky2008optical,falkovsky2007space,falkovsky2008opticalconference,falkovsky2007optical,gusynin2007sum,kargarian2015theory} and follow the prescription outlined in Ref.~\cite{jo-2017} to compute $g_0$ and $I$, i.e., we postulate the presence of an induced Kerr nonlinearity in the slab and determine $g_0$ and $I$ by demanding that the corresponding Helmholtz equation respectively develops a linear and a nonlinear spectral singularity.

\section{Determination of the threshold gain and output intensity}
\label{S2Ch3}

We model the electromagnetic properties of the thin layers coating the slab by demanding that its conductivity has the form: $\sigma=\sum_{j=1}^{2}\sigma_{j}\,\delta(z-z_{j})$,
where $ \sigma_{j} $ with $ j=1,2 $ are constant coefficients representing the contributions of the $j$-th layer, and $\delta(\cdot)$ stands for the Dirac delta function. We also recall that the conductivity determines the current density according to
    \[\vec J:=\sigma \vec{E} =\sum_{j=1}^{2}\sigma_{j}\delta(z-z_{j})\vec{E}(\vec{r}).\]

For TE waves, the electric field $\vec{E}(\vec{r})$ takes the form: $\vec{E}(\vec{r})=e^{ik_{x}x}\cE(z)\vec{e}_{y}$, where $ \cE(z) $ is a complex amplitude. Using Maxwell's equations, we can show that it satisfies the nonlinear Helmholtz equation,
    \bea
    \cE''(z)+k^{2}[\hat{\epsilon}(z,\cE)-\sin^{2}\theta]\cE(z)+
    ikZ_{0}\sum_{j=1}^{2}{J}_{j}(z)=0,
    \label{NL Helmholtz eqs.}
    \eea
where $\hat{\epsilon}$ stands for the nonlinear relative
permittivity of the slab \cite{boyd2003nonlinear}, namely
    \bea
    \hat{\epsilon}(z,\cE)&:=&\left\{\begin{array}{ccc}
    \bn^{2}+\varsigma\vert\cE(z)\vert^{2} &{\rm for}&z\in[0,L],\\
    1&{\rm for}&z\notin[0,L],
    \end{array}\right.\nn
    \eea
$\bn$ denotes the complex refractive index of the slab, $\varsigma$ is the Kerr coefficient,  $Z_{0}:=\sqrt{\mu_0/\epsilon_0}$, $\epsilon_{0}$, and $ \mu_{0} $ are respectively the impedance, permittivity, and permeability of the vacuum, and
    \be
    {J}_{j}(z):=\sigma_{j}\delta(z-z_{j})\cE(z).
    \nn
    \ee

To simplify the calculations, we introduce:
\begin{align}
    &\hat{z}:=\dfrac{z}{L}, & &\fK := kL\cos\theta,
    \label{parameter1}\\
    &\gamma:=-\varsigma k^{2}L^{2}, &
    &\tilde\bn :=\sec\theta\sqrt{\bn^2-\sin^2\theta},
    \label{parameter2}\\
    &\psi(\hat{z}):=\cE(\hat{z}L), &
    &\hat{\sigma}_{j}:=-i\fK \,Z_{0}\sec\theta\,\sigma_{j}.
    \label{parameter3}
\end{align}
$ \hat{\sigma}_j$, which marks the scaled conductivity parameters, depends on the scaled wavenumber $\fK$ and a set of physical parameters, $\vartheta_{1}, \vartheta_{2}, \cdots,\vartheta_{N}$, that describe the 2D material used for coating the slab;
    \be
    \hat{\sigma}_j=
    \hat{\sigma}_j(\fK,\vartheta_{1}, \vartheta_{2}, \cdots,\vartheta_{N}).
    \label{hat-sigma=}
    \ee

In view of (\ref{parameter1}) -- (\ref{parameter3}), we can identify (\ref{NL Helmholtz eqs.}) with the nonlinear Schr\"odinger equation,
    \bea
    -\psi''(\hat{z})+\chi(\hat{z})[v(\hat{z})+
    \gamma\vert\psi(\hat{z})\vert^{2}]\psi(\hat{z})=\fK^{2}\psi(\hat{z}),
    \label{NSE}
    \eea
where
    \bea
    v(\hat{z})&:=&\fK^{2}(1-\tilde\bn^{2})
    +\sum_{j=1}^{2}\hat\sigma_{j}\,\delta(\hat z-\hat z_{j}),\nn\\
    \chi(\hat{z})&:=&\left\{\begin{array}{ccc}
    1 ~~~{\rm for}~~~\hat{z}\in[0,1],\\
    0~~~{\rm for}~~~\hat{z}\notin[0,1].
    \end{array}\right.
    \nn
    \eea
It is not difficult to show that the standard boundary conditions fulfilled by electric and magnetic fields \cite{jackson1999classical} are equivalent to the requirement that $ \psi(\hat{z}) $ be continuous at $\hat{z}=\{0,1\}$ and
    \be
    \begin{aligned}
    &\psi'(0^{+})=\psi'(0^{-})+\hat{\sigma}_{1}\psi(0),\\
    &\psi'(1^{+})=\psi'(1^{-})+\hat{\sigma}_{2}\psi(1),
    \end{aligned}
    \label{TEmatching condition}
    \ee
where $\psi'(\hat z_0^{+})$ and $\psi'(\hat z_0^{-})$ respectively stand for the right and left limits of $\psi'(\hat z)$ as $\hat z\to\hat z_0$.

For TM waves, the magnetic field is given by
    \bea
    \vec{H}(\vec{r})=e^{ik_{x}x}\cH(z)\vec{e}_{y}\label{TM H},
    \nn
    \eea
where $ \cH(z)$ is a complex amplitude. Ref.~\cite{jo-2017} gives the magnetic analog of the nonlinear Helmholtz equation (\ref{NL Helmholtz eqs.}) and establishes its equivalence to the following nonlinear Schr\"{o}dinger equations \cite{jo-2017}.
    \be
    -\phi''(\hat{z})+\chi(\hat z)\left[\fK^{2}(1-
    \tilde{\bn}^{2})\phi(\hat{z})+\gamma F_{0}(\hat{z})\right]=\fK^{2}\phi(\hat{z}),
    \label{NSE for TM}
    \ee
where $\phi(\hat{z}):=\cH(\hat{z}L)$ and
    \bea
    F_{0}(\hat{z})&:=&\sec^{2}\theta f_{0}(\hat{z})\phi(\hat{z})
    -\dfrac{\cos^{2}\theta}{\bn^{2}\fK^{2}}f_{0}'(\hat{z})\phi'(\hat{z}),
    \nn
    \\
    f_{0}(\hat{z})&:=&\dfrac{Z_{0}^{2}}{\fK^{2}\vert \bn\vert^{4}}
    [\cos^{2}\theta \vert\phi'(\hat{z})\vert^{2}+\sin^{2}\theta\fK^{2}\vert
    \phi(\hat{z})\vert^{2}].\nn
    \eea
Furthermore, the matching conditions satisfied by $\phi(\hat{z})$ at $ \hat{z}=\{0,1\}$ are given by
\begin{equation}
    \begin{aligned}
    &\phi'(0^{+})=\hat\bn_{1}^{2}\phi'(0^{-}),\quad
    \phi(0^{+})-\phi(0^{-})
    =-\dfrac{\hat \sigma_{1}\cos^{2}\theta}{\fK^{2}}\phi'(0^{+}),\\
    &\hat\bn_{2}^{2}\phi'(1^{+})=\phi'(1^{-}),\quad \phi(1^{+})-\phi(1^{-})
    =-\dfrac{\hat\sigma_{2}\cos^{2}\theta}{\fK^{2}}\phi'(1^{+}),
    \end{aligned}\label{TMmatching condition}
\end{equation}
where
    \[\hat\bn_{j}^{2}:=\bn^{2}-\cos^{2}\theta\fK^{-2}f_{0}(z_{j})\gamma+\cO(\gamma^{2}),\]
and $\cO(\gamma^{2})$ labels the quadratic and higher order terms in powers of $\gamma$.

Next, we consider the following solution of the nonlinear Schr\"{o}dinger equations for the TE and TM waves that describes the scattering of a left incident wave.
    \bea
    \Psi(x)=
    \begin{cases}
    \dfrac{\cG^{E/M}_{+}e^{i\fK \hat{z}}-\cG^{E/M}_{-}e^{-i\fK\hat{z}}}{2i\fK}
    & \hat{z} < 0,\\
    \xi_{E/M}(\hat{z}) & 0 \leq \hat{z}  \leq 1,\\[6pt]
    N_{+}e^{i\fK \hat{z}} & \hat{z} > 1,
    \end{cases}
    \label{Sol. of NSE}
    \eea
where $ N_{+} $ is the complex amplitude of the transmitted wave, $\xi_{E}(\hat{z})$ and $\xi_{M}(\hat{z})$ are respectively the solutions of (\ref{NSE}) and (\ref{NSE for TM}) on $[0,1]$ that fulfill the initial conditions:
    \begin{align}
    &\xi_{E}(1)=N_{+}e^{i\fK},
    &&\xi_{E}'(1)=(i\fK-\hat{\sigma}_{2})N_{+}e^{i\fK},
    \nn
    \\
    &\xi_{M}(1)=N_{+}e^{i\fK}(1+\dfrac{i\hat{\sigma}_{2}
    \cos^{2}\theta}{\fK}),
    &&\xi_{M}'(1)=i\fK\,\hat{\bn}_{2}^{2}N_{+}e^{i\fK},
    \nn
    \end{align}
and $\cG^{E/M}_{\pm}$ are coefficients that we can determine by imposing the relevant matching conditions at  $\hat{z}=0$. In view of  (\ref{TEmatching condition}) and  (\ref{TMmatching condition}), we have
    \bea
    \cG^{E}_{\pm}&=&\xi_{E}'(0)\pm(i\fK\mp \hat{\sigma}_{1})\xi_{E}(0),\nn\\
    \cG^{M}_{\pm}&=&\pm i\fK\,\xi_{M}(0)+
    \left(\dfrac{\fK\pm i\hat{\sigma}_{1}
    \cos^{2}\theta\hat\bn_{1}^{2}}{\fK\,\hat\bn_{1}^{2}} \right) \xi'_{M}(0).\nn
    \eea

According to Eq.~(\ref{Sol. of NSE}), the left reflection and transmission amplitudes are given by,
    \begin{align}
    &R_{E/M}^{l}=-\frac{\cG^{E/M}_{-}}{\cG^{E/M}_{+}},
    &&T_{E/M}^{l}=\dfrac{2i\fK N_{+}}{\cG^{E/M}_{+}}.\nn
    \end{align}
Because nonlinear spectral singularities \cite{prl-2013}  correspond to singularities of $R_{E/M}^{l}$ and $T_{E/M}^{l}$, we can characterize them using the values of $\fK$ such that
    \bea
    \cG^{E/M}_{+}=0.
    \label{NSS condition}
    \eea

For a weak Kerr nonlinearity, where $ \gamma\ll 1$, we can expand the relevant quantities in a power series in $\gamma$ and ignore quadratic and higher order terms in power of $\gamma$, i.e., employ first-order perturbation theory \cite{pra-2013,jo-2017}. In this way, we can express (\ref{NSS condition}) in the form
    \bea
    \cG^{E/M}_{0+}+\gamma\,\cG^{E/M}_{1+}+\cO(\gamma^{2})
    = 0,
    \label{firstorder perturbation}
    \eea
where
    \bea
    &&\cG_{0+}^{E}:=
    \xi_{E0}^{\prime}(0)+(i\fK- \hat{\sigma}_{1})\,\xi_{E0}(0),
    \label{G0E}
    \\
    &&\cG_{1+}^{E}:=\xi_{E1}^{\prime}(0)+(i\fK- \hat{\sigma}_{1})\,
    \xi_{E1}(0),
    \label{G1E}
    \\
    &&\xi_{E0}(\hat{z}):=\dfrac{N_{+}e^{i\fK}}{2i\fK\,\tilde{\bn}}
    \left[e^{{i\fK\tilde{\bn}(\hat{z}-1)}}\alpha_{E+}
    +e^{-{i\fK \tilde\bn(\hat{z}-1)}}\alpha_{E-}\right],
    \nn\\
    &&\xi_{E1}(\hat{z}):=\int_{1}^{\hat{z}}\dfrac{
    \sin[\fK\tilde{\bn}(\hat{z}-\hat{z}')]}{\fK\,\tilde{\bn}}
    \left|\xi_{E0}(\hat{z}')\right|^{2}
    \xi_{E0}(\hat{z}')d\hat{z}',
    \nn\\
    &&\alpha_{E\pm}:=(\tilde\bn\pm1)\pm \dfrac{i}{\fK}
    \,\hat\sigma_{2},\nn
    \eea
and
    \bea
    &&\cG_{0+}^{M}:=i\fK\bn^{2}\xi_{M0}(0)
    +\left(1+\dfrac{i\hat{\sigma}_{1}\bn^{2}\cos^{2}\theta}{\fK}\right)
    \xi_{M0}^{\prime}(0),
    \label{G0M}\\
    &&\cG_{1+}^{M}:=
    -\dfrac{i\cos^{2}\theta}{\fK}f_{0}(0)\left[ \xi_{M0}(0)
    +\dfrac{\hat{\sigma}_{1}\cos^{2}\theta}{\fK^{2}}\xi_{M0}^{'}(0)\right]
    \nonumber\\
    &&~~~~~+\left(1+\dfrac{i\hat{\sigma}_{1}\bn^{2}}{\fK}
    \cos^{2}\theta\right)\tilde \xi_{M1}^{\prime}(0)+
    i\fK\bn^{2} \tilde \xi_{M1}(0),
    \label{G1M}\\
    &&\xi_{M0}(\hat{z}):=\dfrac{N_{+}e^{i\fK}}{2i\fK\tilde{\bn}}
    \left[e^{{i\fK\tilde{\bn}(\hat{z}-1)}}\alpha_{M+}
    +e^{-{i\fK\tilde{\bn}(\hat{z}-1)}}\alpha_{M-}\right],
    \nn\\
    &&\tilde \xi_{M1}(\hat{z}):=
    -\dfrac{iN_{+}e^{i\fK}\cos^{2}\theta}{\tilde{\bn}
    \fK^{2}}f_{0}(1)\sin[\fK\tilde{\bn}(\hat{z}-1)]+\xi_{M1}(\hat{z}),
    \nn\\
    &&\xi_{M1}(\hat{z}):=\int_{1}^{\hat{z}}\dfrac{\sin[\fK\tilde{\bn}
    (\hat{z}-\hat{z}')]}{\fK\tilde{\bn}}F_{0}(\hat{z}')d\hat{z}',
    \nn\\
    &&\alpha_{M\pm}:=\tilde\bn\pm \bn_{0}^{2}+\dfrac{i\tilde\bn\cos^{2}\theta\hat\sigma_{2}}{\fK}.
    \nn
    \eea

A linear spectral singularity arises when the first term on the left-hand side of (\ref{firstorder perturbation}) vanishes. Denoting the corresponding values of $\bn$, $\fK$, and $\hat{\sigma}_{j}$ respectively by $\bn_{0}$, $\fK_{0}$, and $\hat{\sigma}_{0j}$, we can express this condition as
    \bea
    \cG^{E/M}_{0+}(\fK_{0},\bn_{0},
    \hat{\sigma}_{01},\hat{\sigma}_{02})=0,
    \label{LSS condition}
    \eea
where we have made the $\fK$-, $\bn$-, and $\hat{\sigma}_{j}$-dependence of  $\cG^{E/M}_{0+}$ explicit. Note that in view of (\ref{hat-sigma=}), $\hat{\sigma}_{0j}$ are determined by certain values $\vartheta_{0\ell}$ of the physical parameters $\vartheta_\ell$  of the 2D material coating the faces of the slab. Introducing the abbreviation,
    \begin{align}
    &\bvtheta:=(\vartheta_{1},\vartheta_{2},\cdots,\vartheta_{N}),
    &&\bvtheta_{0}:=(\vartheta_{01},\vartheta_{02},\cdots,\vartheta_{0N}),\nn
    \end{align}
we can view $\cG^{E/M}_\pm$ as functions of $\fK,\bn$, and $\bvtheta$; $\cG^{E/M}_\pm=\cG^{E/M}_\pm(\fK,\bn,\bvtheta)$, and write the condition (\ref{LSS condition}) for the presence of a linear spectral singularities in the form:
    \be
    \cG^{E/M}_{0+}(\fK_{0},\bn_{0},\bvtheta_{0})=0.
    \label{LSS condition2}
    \ee

To explore the physical content of (\ref{LSS condition}) and (\ref{LSS condition2}), first we show their equivalence to
    \bea
    e^{2i\fK_{0}\tilde\bn_{0}}=
    \dfrac{\alpha^{E/M}_{0+}\beta^{E/M}_{0+}}{\alpha^{E/M}_{0-}\beta^{E/M}_{0-}},
    \label{linear SS}
    \eea
where
    \begin{align}
    &\alpha^{E}_{0\pm}:=\tilde\bn_{0}\pm1 \pm\dfrac{i}{\fK}\hat\sigma_{02},
    \label{alpha0E}
    &&\alpha^{M}_{0\pm}:=\tilde\bn_{0}\pm \bn_{0}^{2}+
    \dfrac{i\tilde\bn_{0}\cos^{2}\theta\hat\sigma_{02}}{\fK_{0}},
    \\
    &\beta^{E}_{0\pm}:=\tilde\bn_{0}\pm 1\pm\dfrac{i}{\fK} \hat\sigma_{01},
    \label{alpha0M}
    &&\beta^{M}_{0\pm}:=\tilde\bn_{0}\pm \bn_{0}^{2}+
    \dfrac{i\bn_{0}^{2}\tilde\bn_{0}\cos^{2}\theta\hat
    \sigma_{01}}{\fK_{0}}.
    \end{align}
We then recall the relation,
    \bea
    g=-2k\kappa,
    \label{relation for gain coeff.}
    \eea
between the gain coefficient $g$ and the imaginary part $\kappa$ of the refractive index $\bn$, \cite{silfvast}. The threshold gain coefficient $g_0$ is the value of $g$ at which the optical system acquires a linear spectral singularity \cite{pra-2011a}, i.e., (\ref{linear SS}) holds. According to (\ref{relation for gain coeff.}), it is given by
    \be
    g_{0}=-2k_{0}\kappa_{0},
    \label{g-zero=}
    \ee
where $k_0:=\fK_0/L\cos\theta$ is the corresponding wavenumber, and $\kappa_0:=\IM(\bn_0)$. Equating the modulus and phase angle of the left- and right-hand sides of (\ref{linear SS}) and making use of (\ref{g-zero=}), we obtain the following formulas for the threshold gain $g^{E/M}_{0}$ for lasing in the TE/TM modes of the slab and the corresponding scaled wavenumber $\fK_0$.
    \bea
    &&g^{E/M}_{0}=\dfrac{\IM(\bn_{0})}{L\cos\theta\, \IM(\tilde{\bn}_{0})}
    \ln\left| \frac{\alpha^{E/M}_{0+}\beta^{E/M}_{0+}}{
    \alpha^{E/M}_{0-}\beta^{E/M}_{0-}}\right|,
    \label{threshold gain} \\
    &&\fK_{0}=\dfrac{\pi m-\varphi_{0}}{\RE(\tilde{\bn}_{0})},
    \label{threshold K}
    \eea
where $m$ is a positive integer, $\varphi_{0}$ is the phase angle of $\alpha^{E/M}_{0+}\beta^{E/M}_{0+}/\alpha^{E/M}_{0-}\beta^{E/M}_{0-}$, and `$\RE$' and `$\IM$' respectively stand for the real and imaginary part of their argument.

To characterize nonlinear spectral singularities, we solve (\ref{NSS condition}) using first-order perturbation theory with the scaled Kerr coefficient $ \gamma $ playing the role of the perturbation parameter.

Let $ \fK $, $ \bn $ and $ \bvtheta$ be respectively the values of wavenumber, complex refractive index, and the collection of the physical parameters of the coatings that give rise to a nonlinear spectral singularity. Expanding these in powers of $\gamma$ and ignoring quadratic and higher order terms, we have
    \begin{align}
    &\fK=\fK_{0}+   \gamma\fK_{1},
    &&\bn=\bn_{0}+\gamma\bn_{1},
    &&\bvtheta=\bvtheta_{0}+\gamma \bvtheta_{1},
    \label{nonlinear expansion of parameters}
    \end{align}
where $\fK_{1}$, $\bn_1$, and $\bvtheta_{1}:=(\vartheta_{11},\vartheta_{12},\cdots,\vartheta_{1N})$ are $\gamma$-independent parameters signifying the contribution of the nonlinearity. We also use $ \eta_{1} $ and $ \kappa_{1} $ to label the real and imaginary parts of $\bn_1$, so that $ \bn_{1}:=\eta_{1}+i\kappa_{1} $, and recall that in view of (\ref{relation for gain coeff.}) and (\ref{nonlinear expansion of parameters}),
    \bea
    g=g_{0}\left[1+\gamma\left(\dfrac{\fK_{1}}{\fK_{0}}+
    \dfrac{\kappa_{1}}{\kappa_{0}} \right)  \right],
    \label{relation for gain}
    \eea
where we have dropped the quadratic and higher order terms in $\gamma$,  \cite{pra-2013}.

 In light of (\ref{LSS condition}) and (\ref{nonlinear expansion of parameters}), we can satisfy (\ref{NSS condition}) up to and including the first-order terms in $\gamma$ provided that
    \bea
    &&
    \cG_{1+}^{E/M}(\fK_{0},\bn_{0},\bvtheta_{0})+
    \partial_{\fK_{0}}\cG_{0+}^{E/M}(\fK_{0},\bn_{0},\bvtheta_{0})\,\fK_{1}+\nn\\
    &&\partial_{\bn_{0}}\cG_{0+}^{E/M}(\fK_{0},\bn_{0},\bvtheta_{0})\,\bn_{1}
    +\sum_{\ell=1}^{N}
    \partial_{\vartheta_{0\ell}}\cG_{+}^{E/M(0)}(\fK_{0},\bn_{0},\bvtheta_{0})
    \vartheta_{1\ell}
    =0.\nn
    \eea
Solving this equation for $\fK_1$ and using (\ref{G0E}) --  (\ref{G1M}), we have been able to show that
    \bea
    \fK_{1}=\fa\,I+\fb\,\bn_{1}+\sum_{\ell=1}^{N}\fc_{\ell}\,\vartheta_{1\ell},
    \label{K1 relation}
    \eea
where, $ I=\vert N_{+}\vert^{2}/2 $ is the time-averaged intensity of the emitted wave from the right face of the slab (i.e., the plane $z=L$), and $\fa$, $\fb$ and $\fc_{\ell}$ are the complex coefficients given by,
    \bea
    &&\fa:=\dfrac{-2\cG_{1+}(\fK_{0},\bn_{0},\bvtheta_{0})}{\vert N_{+}
    \vert^{2}\partial_{\fK_{0}}\cG_{0+}(\fK_{0},\bn_{0},\bvtheta_{0})},\nn\\[6pt]
    &&\fb:=\dfrac{-\partial_{\bn_{0}}\cG_{0+}(\fK_{0},\bn_{0},
    \bvtheta_{0})}{\vert N_{+}\vert^{2}\partial_{\fK_{0}}
    \cG_{0+}(\fK_{0},\bn_{0},\bvtheta_{0})},\nn\\[6pt]
    &&\fc_{\ell}:=\dfrac{-\partial_{\vartheta_{0\ell}}
    \cG_{0+}(\fK_{0},\bn_{0},\bvtheta_{0})}{
    \vert N_{+}\vert^{2}\partial_{\fK_{0}}
    \cG_{0+}(\fK_{0},\bn_{0},\bvtheta_{0})}.\nn
    \eea

Next, we introduce
     \begin{align}
    & \fa_{r}:=\RE(\fa),
    && \fb_{r}:=\RE(\fb),
    && {\fc}_{\ell r}:=\RE(\fc_{\ell}),\nn\\
    & \fa_{i}:=\IM(\fa),
    && \fb_i:=\IM(\fb),
    && {\fc}_{\ell i}:=\IM(\fc_{\ell}).\nn
    \end{align}
Noting that $\fK_1$ is real, the imaginary part of the right-hand side of
(\ref{K1 relation}) vanishes identically. We have used this observation together with (\ref{relation for gain}) and (\ref{K1 relation}) to derive the following expression for the time-averaged intensity of the emitted wave.
    \bea
    I=\dfrac{g-g_{0}}{\varsigma \, g_{0}}\:\hat I_{g}+
    \dfrac{\eta-\eta_{0}}{\varsigma\, \eta_{0}}\:\hat I_{\eta}
    +\sum_{\ell=1}^{N}\dfrac{\vartheta_{\ell}-\vartheta_{0\ell}}{\varsigma \,\vartheta_{0\ell}}\:\hat I_{\vartheta_{\ell}},
    \label{intensity-relation}
    \eea
where $\eta_{0}$ is the real part of $\bn_0$, so that $\bn_0=\eta_0+i\kappa_0$, and
    \bea
    && \hat I_{g}:=-
    \dfrac{\fb_{r}\kappa_{0}\cos^{2}\theta}{\fK_{0}[\fa_{i}(\fb_{i}\kappa_{0}-\fK_{0})+\fa_{r}\fb_{r}\kappa_{0}]},
    \nonumber\\[6pt]
    && \hat I_{\eta}:=\dfrac{\eta_{0}\cos^{2}\theta}{\fK_{0}^{2}}\left[\dfrac{\fb_{i}(\fb_{i}\kappa_{0}-\fK_{0})+\fb_{r}^{2}\kappa_{0}}{\fa_{i}(\fb_{i}\kappa_{0}-\fK_{0})+\fa_{r}\fb_{r}\kappa_{0}} \right],
    \nonumber\\[6pt]
    &&\hat I_{\vartheta_{\ell}}:=\dfrac{\vartheta_{0\ell}\cos^{2}\theta}{\fK_{0}^{2}}\left[\dfrac{{\fc}_{\ell i}(\fb_{i}\kappa_{0}-\fK_{0})+\fb_{r}{\fc}_{\ell r}\kappa_{0}}{\fa_{i}(\fb_{i}\kappa_{0}-\fK_{0})+\fa_{r}\fb_{r}\kappa_{0}} \right].\nn
    \eea

Eq.~(\ref{intensity-relation}) is the main result of this article. In order for the slab to emit laser light from its right face ($z=L$), the right-hand side of (\ref{intensity-relation}) should take a positive value. In the absence of the coatings $\hat I_{\vartheta_{\ell}}=0$ and one can easily arrange $\eta$ such that this condition is fulfilled for $g>g_0$ except for the TM modes with incidence (emission) angle $\theta$ exceeding the Brewster's angle. This is an indication that lasing in these modes of the slab are forbidden \cite{jo-2017}. In the following section, we explore the consequences of Eq.~(\ref{intensity-relation}) for a slab with Graphene or WSM coatings on one or both of its faces. This requires a detailed examination of the intensity slopes $\hat I_{g}$, $\hat I_{\eta}$, and $\hat I_{\vartheta_\ell}$. Alternatively, we can explore the structure of $\hat I_{g}$ and the following ratios of intensity slopes.
    \begin{align}
    &\cX_\eta:=\frac{\hat I_\eta}{\hat I_g}=\eta_0\left[\frac{\fb_i}{\fb_r}\left(\frac{1}{\kappa_0}-\frac{\fb_i}{\fK_0}\right)-\frac{\fb_r}{\fK_0}\right],
    \label{cX-eta}\\
    &\cX_{\vartheta_\ell}:=\frac{\hat I_{\vartheta_\ell}}{\hat I_g}=
    \vartheta_{0\ell}\left[\frac{\fc_{\ell i}}{\fb_r}\left(\frac{1}{\kappa_0}-\frac{\fb_i}{\fK_0}\right)-\frac{\fc_{\ell i}}{\fK_0}\right].
    \label{cX-theta}
    \end{align}
In terms of these, (\ref{intensity-relation}) takes the form,
    \bea
        I&=&\frac{\hat I_g}{\varsigma}\left[
        \dfrac{g-g_{0}}{g_{0}}+
    \cX_\eta\left(\dfrac{\eta-\eta_{0}}{\eta_{0}}\right)+
        \sum_{\ell=1}^{N} \cX_{\vartheta_\ell}
    \left(\dfrac{\vartheta_{\ell}-\vartheta_{0\ell}}{\vartheta_{0\ell}}\right)\right],\nn\\
    &=&I_0\left(1+\cX_\eta\,\delta\eta+\sum_{\ell=1}^{N} \cX_{\vartheta_\ell}\,\delta\vartheta_{\ell}\right),
    \label{intensity-relation-n}
    \eea
    where
        \begin{align*}
    &I_0:=\frac{\hat I_g(g-g_{0})}{\varsigma\,g_{0}},
    &&\delta\eta:=\frac{\eta/\eta_0-1}{g/g_0-1},
    &&\delta\vartheta_{\ell}\:=\frac{\vartheta_{\ell}/\vartheta_{0\ell}-1}{g/g_0-1}.
    \end{align*}

According to (\ref{intensity-relation-n}), $I_0$ is the intensity of the emitted wave from the right-hand face of the slab when $\eta=\eta_0$ and $\vartheta_\ell=\vartheta_{0\ell}$. Depending on the values of $\cX_\eta$ and $\cX_{\vartheta_\ell}$, we can in principle adjust $\eta$ and $\vartheta_\ell$ so that $I>I_0$. This means that we can boost the intensity of the emitted wave without increasing the gain coefficient of the slab. In practice, it should be easier and more convenient to do this by tuning the parameters $\vartheta_\ell$ of the conductivity of the coatings, for it would not involve making changes in the active material constituting the slab.

\section{Physical realization and implications of the results}

Coating the faces of a homogeneous slab laser by a thin layer of graphene or WSM introduces the tunable parameters $\vartheta_\ell$ into the expression for the threshold gain \cite{sarisaman2018pt}. In the preceding section we have shown that the same happens for the laser output intensity. To identify the nature of these parameters and study their effects on the properties of a concrete slab laser, we first examine the conductivities of the graphene and 2D WSM.

The conductivity of graphene \cite{falkovsky2008optical}, which we denote by $\sigma_g$, is the sum of the intraband and interband contributions; $\sigma_{g}=\sigma_{\rm intra}+ \sigma_{\rm inter}$, where
    \begin{flalign}
    &\sigma_{\rm intra}:=
    \frac{ie^{2}\sE \ln[2\cosh(\mu/\sE)]}{\pi\hbar^{2}(\omega+i\Gamma)},
    \label{gintra}\\
    &\sigma_{\rm inter}:=\dfrac{e^{2}}{4\pi\hbar}\left[\dfrac{\pi}{2}
    +\tan^{-1}\left(\frac{\nu_{-}}{\sE}\right)
    -\dfrac{i}{2}\ln\left(\dfrac{\nu_{+}^{2}}{\nu_{-}^{2}+\sE^{2}}\right)\right],
    \label{ginter}
    \end{flalign}
$-e$ is the electron charge, $\sE:=2k_{B}T$, $k_{B}$ is the Boltzmann constant,  $T$ is the temperature, $\mu$ is the chemical potential, $\Gamma$ is the scattering rate of charge carriers, $\nu_{\pm}:=\hbar\omega\pm2\mu$, and $\hbar\omega $ is the photon energy.
In pure graphene the chemical potential is zero, and the intraband conductivity (\ref{gintra}) is proportional to the temperature. In general, the electron density $ \mathit{n}_{0}$ is related to the temperature and the chemical potential according to $\mathit{n}_{0}=(2/\pi \hbar\mathit{v}_{0})\int_{0}^{+\infty}\varepsilon[\mathit{f}(\varepsilon-\mu)-\mathit{f}(\varepsilon+\mu)]d\varepsilon$, where $\mathit{f}(\varepsilon)$ is the Fermi function which involves temperature, and $ \mathit{v}_{0}=10^{6} \rm ms^{-1}$ is a constant velocity parameter \cite{falkovsky2008optical}. For a fixed value of the electron density one can use the above formula to determine the temperature-dependence of the chemical potential \cite{falkovsky2007optical}. 

    Next, consider coating the faces of our slab by WSM thin films whose thickness $d$ is much smaller than the wavelength $\lambda$ but much larger than the lattice constant $a$, \cite{yang2011, chu2019, yang2014, burkov2011}. Then as discussed in Ref.~\cite{kargarian2015theory}, the light hits the surface which does not support Fermi-arc states and nodes separated along the $z$-direction. Therefore, their conductivity which we denote by $\sigma_w$
takes the form,
    \bea
    \sigma_{w}\approx
    \dfrac{ie^{2}\ln(2b\lambda)}{2\pi^{2}\hbar}
    =\dfrac{ie^{2}\ln(4\pi\lambda/b')}{2\pi^{2}\hbar},
    \label{2D WSM Conductivity1}
    \eea
where 
$b$ is the measure of the separation between Weyl nodes 
in the $k_{z}$-space, $ b':=2\pi/b $, and we use $\approx $ to imply that the real part of $\sigma_{w}$ is negligibly small compared to its imaginary part.\footnote{Because $a\ll d\ll\lambda$, the cut-off condition for the integral (40) of Ref.~\cite{kargarian2015theory} is satisfied and, in contrast to the case of graphene the interband contributions which give rise to the real part of the surface conductivity negligible. This follows from the use of Kubo formalism for the calculation of the WSM conductivity which involves taking into account and summing up the contribution of individual surface states to the surface currents. In the case we consider, it turned out that the interband transitions do not contribute \cite{kargarian2015theory}.}

In view of Eqs.~(\ref{gintra}), (\ref{ginter}) and (\ref{2D WSM Conductivity1}), we identify the tunable parameters $\vartheta_\ell$ for the Graphene and WSM\footnote{There are various theoretically proposed candidates for a WSM in the literature. Among these TaAs, TaP, NbAs, and NbP have been experimentally observed~\cite{yan2017}. In general, our results apply to any materials that exhibit the WSM phase with surface conductivity given by Eq.~(\ref{2D WSM Conductivity1}).} sheets respectively with the temperature, $T$, and the separation between the Weyl nodes in position space, $b'$. Again, we use the subscript 0 to label the quantities that lead to the emergence of a linear spectral singularity. In particular, $T_0$ and $b'_0$ are the values of $T$ and $b'$ at which the laser attains its threshold gain.

\begin{figure}
    \begin{center}
    \includegraphics[scale=.43]{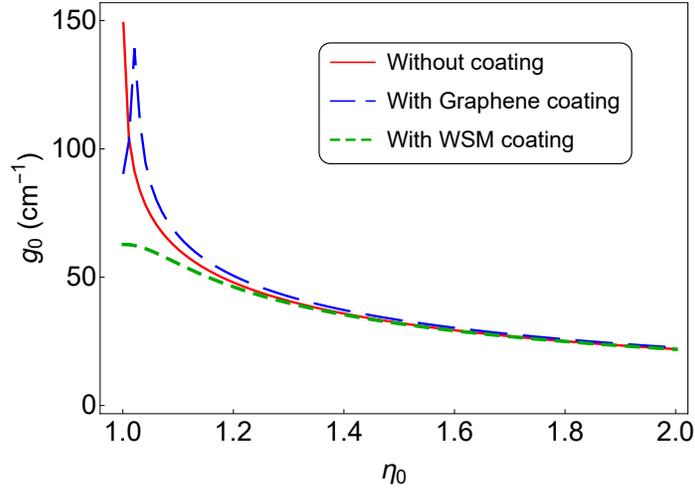}
    \caption{Graphs of the threshold gain coefficient $g_0$ as a function of $\eta_0$ for the normally incident TE mode of wavelength $\lambda=0.5~\mu \rm m $ for $T_0=300~K $, $b'_0=5\times 10^{-4} {\angstrom} $, and $ L=1~{\rm mm}$ in the absence and presence of Graphene or WSM coatings on both faces of the slab.}
    \label{TEgaingraphene}
        \end{center}
    \end{figure}

Fig.~\ref{TEgaingraphene} shows the graphs of the threshold gain (\ref{threshold gain}) as a function of $\eta_{0}$ for the slab with and without a graphene or WSM coating. The effect of the coating becomes negligible for larger values of $\eta_0$ where the internal reflections from the faces of the slab are more pronounced. Our numerical results also show that the reduction of the threshold gain due to coating by a pair of WSM layers diminishes for larger values of $b'_0$.

The behavior of the threshold gain that is depicted in Fig.~\ref{TEgaingraphene} tunes out to be valid for lasing in the oblique TE modes of the slab as well. In general, Graphene or WSM coatings on one or both faces of the slab do not have an appreciable effect on the threshold gain for lasing in TE modes of the slab; the corresponding graphs of $g_0(\theta)$ are almost identical with the one for the slab without a coating \cite{jo-2017}; $g_0$ is a smooth monotonically decreasing function of $\theta$ that tends to zero as $\theta\to 90^\circ$.

\begin{figure*}[htbp!]
    \centering
      \begin{subfigure}{0.45\textwidth}
      \centering
       \captionsetup{justification=centering}
        \includegraphics[scale=.50]{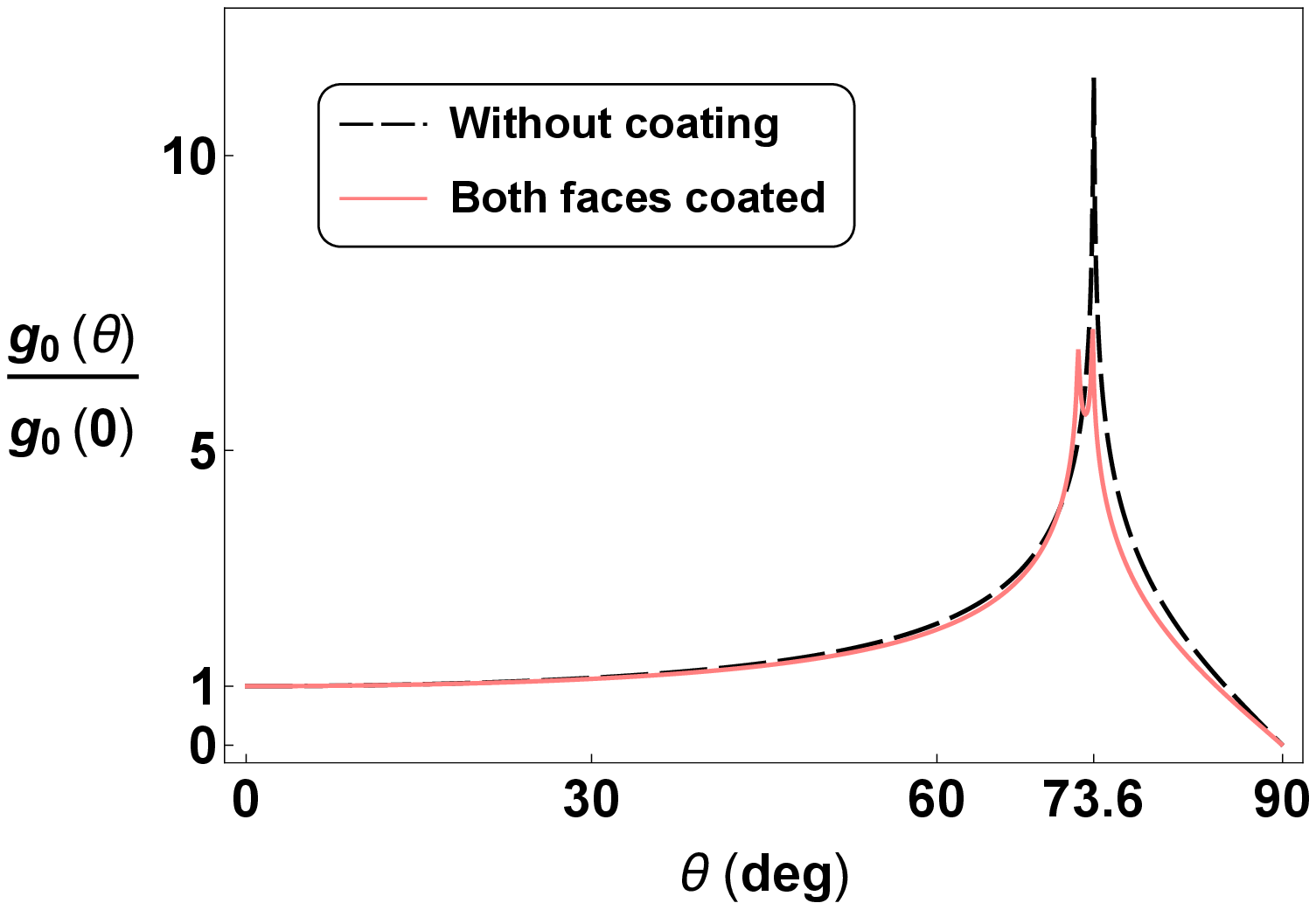}\hspace{0.5cm}
         \caption{}
          \label{fig:NiceImage1}
      \end{subfigure}
      \hfill
      \begin{subfigure}{0.45\textwidth}
      \centering
       \captionsetup{justification=centering}
        \includegraphics[scale=.50]{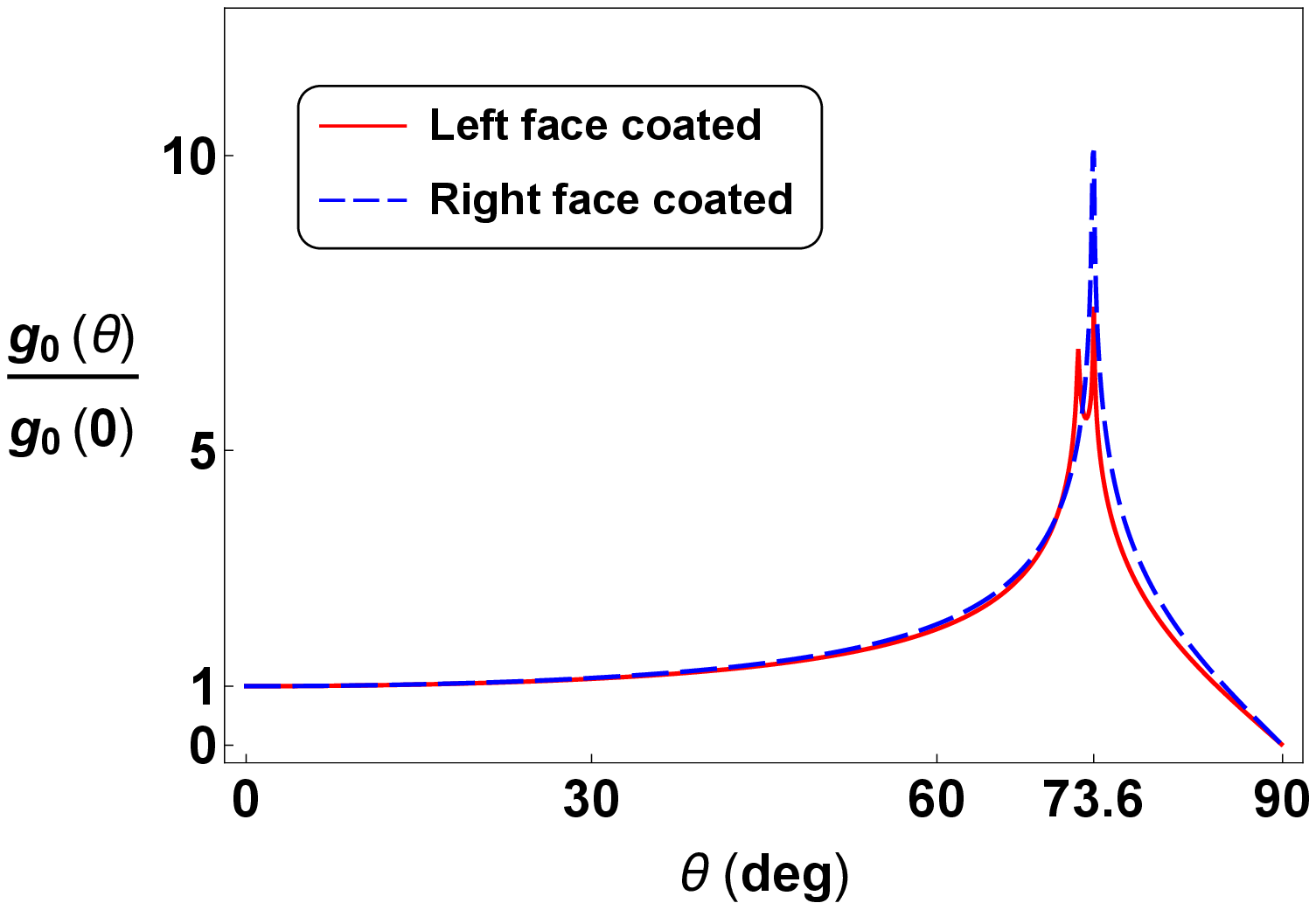}
          \caption{}
          \label{fig:NiceImage2}
      \end{subfigure}\\
           \begin{subfigure}{0.45\textwidth}
            \centering
       \captionsetup{justification=centering}
        \includegraphics[scale=.50]{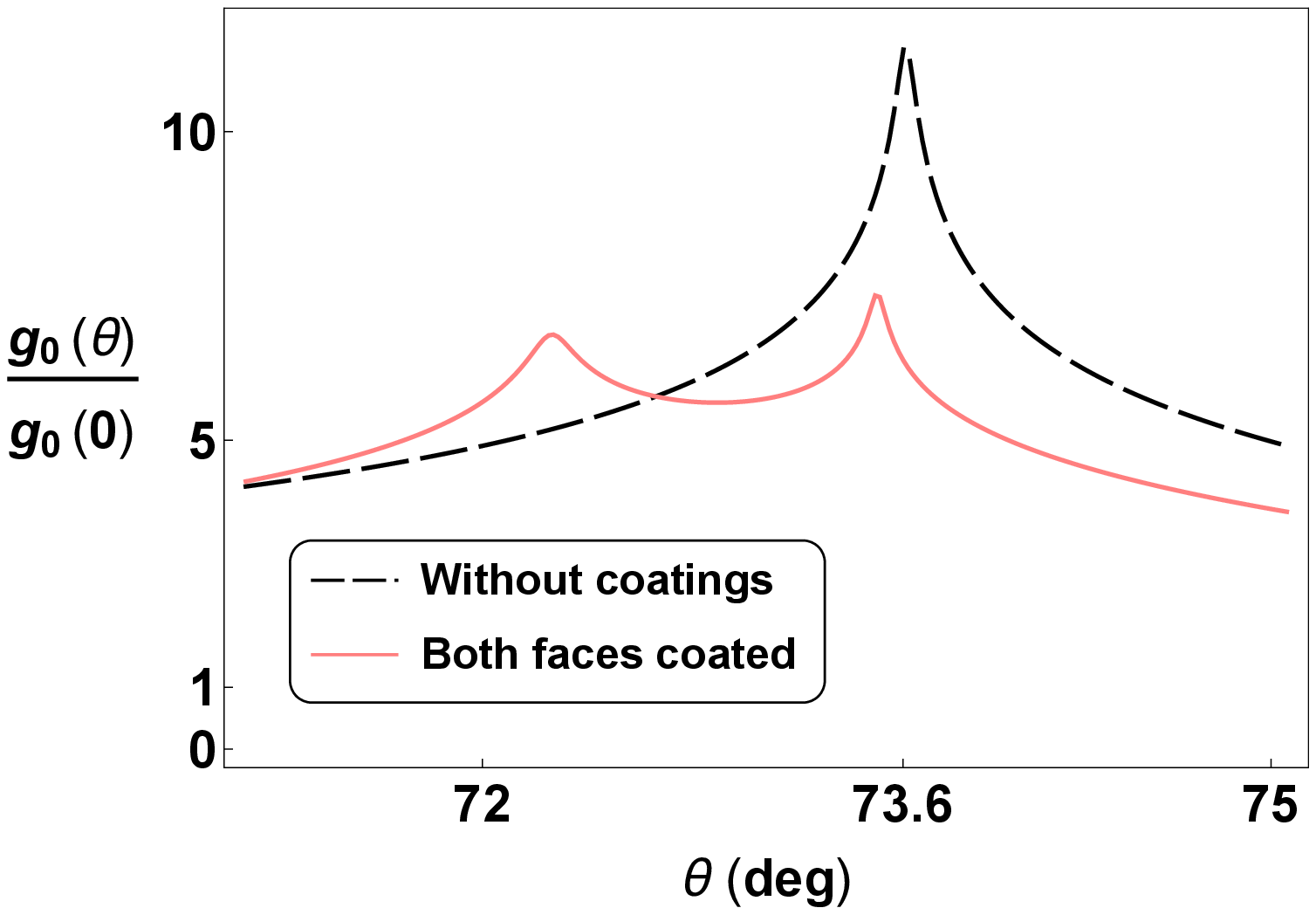}\hspace{0.5cm}
          \caption{}
          \label{fig:NiceImage3}
      \end{subfigure}
      \hfill
      \begin{subfigure}{0.45\textwidth}
      \centering
       \captionsetup{justification=centering}
        \includegraphics[scale=.50]{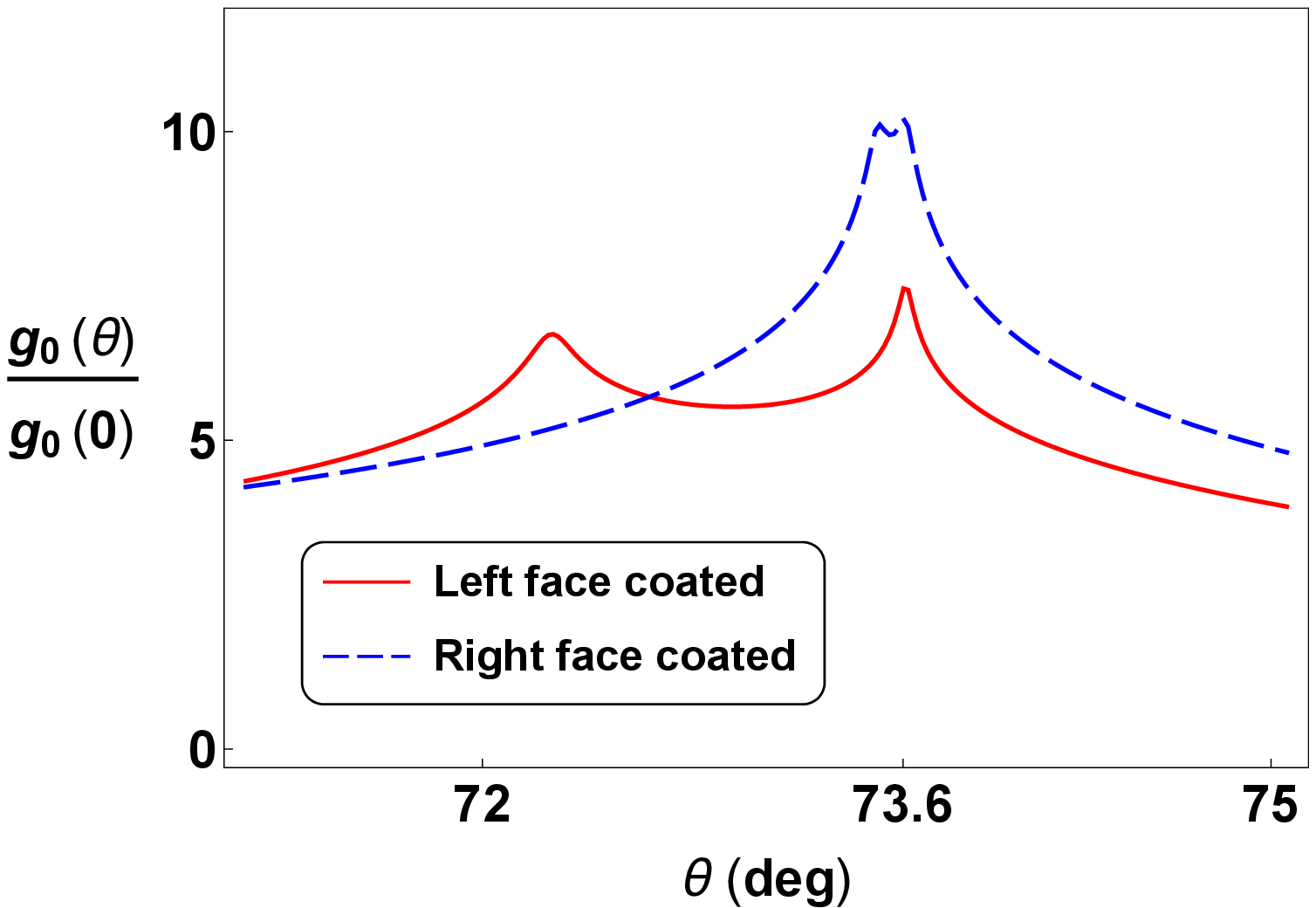}
          \caption{}
          \label{fig:NiceImage4}
      \end{subfigure}
\caption{Graphs of $g_{0}(\theta)/g_{0}(0) $ as a function of $\theta$ for lasing in the TM modes of a slab with and without Graphene coatings for $L=300\,\mu{\rm m}$, $\eta_0=3.4$, $\lambda_0=1.5\,\mu{\rm m}$, and $T_0=300^\circ K$. The graphs in Figs.~\ref{fig:NiceImage3} and \ref{fig:NiceImage4} demonstrate the behavior of $g_{0}(\theta)/g_{0}(0)$ in the vicinity of the Brewster's angle, namely $\theta_b=73.6^\circ$. The values of $g_0(0)$ are listed in Table~\ref{table2}.
\label{TMgaingraphene}%
}
\end{figure*}

The effect of the coatings on the threshold gain is conspicuous for
lasing in the TM modes of the slab. This stems from the difference
between $ \alpha_{-}^{M} $ and $ \beta_{-}^{M} $ as given by
(\ref{alpha0M}). A closer examination of Eq.~(\ref{threshold gain})
shows that its right-hand side has extremely high peaks at a pair of
incidence angles, $\theta_{1}$ and $\theta_{2}$. These respectively
correspond to the values of $\theta$ at which $\beta_{-}^{M}$ and $
\alpha_{-}^{M}$ vanish when we ignore the imaginary part of the
refractive index. In the absence of the coatings, $\theta_1$ and
$\theta_2$ coalesce and coincide with the Brewster angle
$\theta_b:={\rm arctan}(\eta_{0})$, \cite{jo-2017}. In other words,
coating the faces of the slab by Graphene or WSM layers leads to a
splitting of the Brewster angle into a pair of nearby critical
angles. Lasing in the TM modes corresponding to these angles is
effectively impossible, because their threshold gain takes extremely
large values. This is depicted in Fig.~\ref{TMgaingraphene} where we
plot the graphs of the normalized threshold gain as a function of
the incidence (emission) angle $\theta$ for a slab with and without
Graphene coatings.

    \begin{table}
        \begin{center}
        \begin{tabular}{|c|c|}
        \hline
        Coating & $g_0(0)$ (${\rm cm}^{-2}$)\\
    \hline \hline
        None & 40.409 \\
        \hline
        On left face& 46.133 \\
        \hline
        On right face& 40.891 \\
        \hline
        On both faces& 46.616 \\
        \hline
        \end{tabular}
        \vspace{6pt}
        \caption{Values of $g_0(0)$ for the TM modes of the slab with and without Graphene coatings that are considered in Fig.~\ref{TMgaingraphene}.}
        \label{table2}
        \end{center}
        \end{table}
We obtain similar graphs if we coat the same slab using WSM instead of Graphene. The splitting of the Brewster angle is however less noticeable. The same is the case when we coat the slab's right face by a Graphene sheet. Notice that this asymmetry is related to the fact that we consider emission of the laser light from the right face of the slab.

\begin{figure*}
    \begin{center}
    \begin{subfigure}{0.45\textwidth}
      \centering
       \captionsetup{justification=centering}
        \includegraphics[scale=.50]{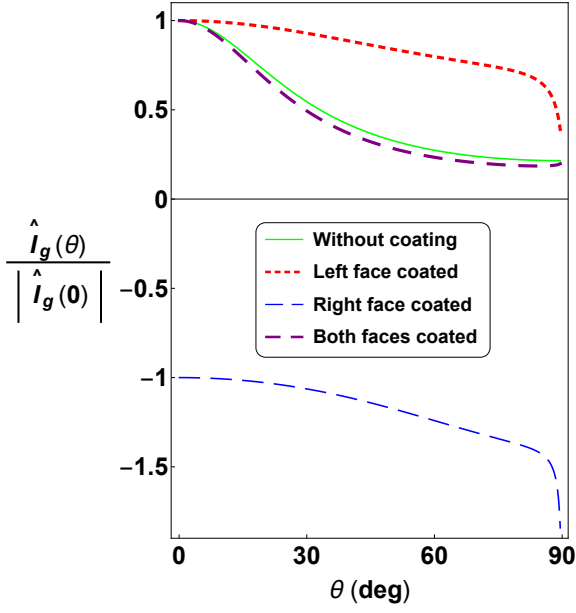}\hspace{0.5cm}
         \caption{}
          \label{fig:NiceImage5}
      \end{subfigure}
      \hfill
      \begin{subfigure}{0.45\textwidth}
      \centering
       \captionsetup{justification=centering}
        \includegraphics[scale=.50]{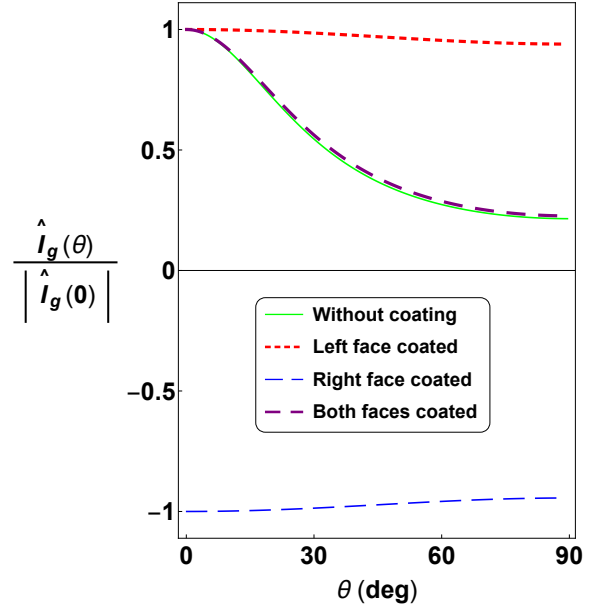}
          \caption{}
          \label{fig:NiceImage6}
      \end{subfigure}
    \caption{Graphs of $ \hat{I}_{g}(\theta)/|\hat{I}_{g}(0)|$ as a function of $\theta$ for lasing in the TE modes of a slab with and without Graphene (Fig.~\ref{fig:NiceImage5} on the left) or WSM (Fig.~\ref{fig:NiceImage6} on the right) coatings for $L=300\,\mu{\rm m}$, $\eta_0=3.4$, $\lambda_0=1.5\,\mu{\rm m}$, $ b'_0=0.0005~{\angstrom} $, and $T_0=300^\circ K$. Values of $\hat{I}_{g}(0)$ for different coating types are listed in Table~\ref{table1}. Coating right face of the slab leads to a negative value of $\hat{I}_{g}(\theta)$.}
    \label{IgTE}
    \end{center}
    \end{figure*}

Fig.~\ref{IgTE} shows plots of the normalized intensity slope $\hat{I}_{g}$ of Eq.~(\ref{intensity-relation}) as a function of $\theta $ for the TE modes of the slab with and without Graphene or WSM coatings. $\hat{I}_{g}$ takes a negative value when the right face of the slab is coated. Therefore, coating the slab's right face prevents emission of laser light in its TE modes from its right-hand face unless we can adjust $\eta, T$, and $b'$ so that they appreciably deviate from their threshold values, $\eta_0, T_0$, and $b'_0$. Moreover, as far as the value of $\hat{I}_{g}$ is concerned coating both faces of the slab has almost the same effect as not coating its faces. 
\begin{table}
        \begin{center}
        \begin{tabular}{|c|c|c|}
        \hline
     Coating & $\hat{I}_{g}(0)$ for TE mode &$\hat{I}_{g}(0)$ for TM mode\\
    \hline \hline
        None & 393.088 & 393.088~$Z_0^{-2}$\\
        \hline
        Graphene on left face & 28.617 & 44.133 $Z_0^{-2}$\\
        \hline
        Graphene on right face & -34.497  & 40.891 $Z_0^{-2}$\\
        \hline
        Graphene on both faces & 473.593 &46.616 $Z_0^{-2}$\\
        \hline
        WSM on left face&  0.259 & -193.124 $Z_0^{-2}$\\
        \hline
        WSM on right face& -0.259 &336.077 $Z_0^{-2}$\\
        \hline
        WSM on both faces& 368.016 &-212.266 $Z_0^{-2}$\\
        \hline
        \end{tabular}
        \vspace{6pt}
        \caption{Values of $\hat{I}_{g}(0)$ for the TE and TM modes
        of the slab considered in Figs.~\ref{IgTE} and \ref{IgTM-WSM}. $Z_0$ is the vacuum impedance.}
        \label{table1}
        \end{center}
        \end{table}

\begin{figure*}
    \begin{center}
    \begin{subfigure}{0.45\textwidth}
      \centering
       \captionsetup{justification=centering}
        \includegraphics[scale=.50]{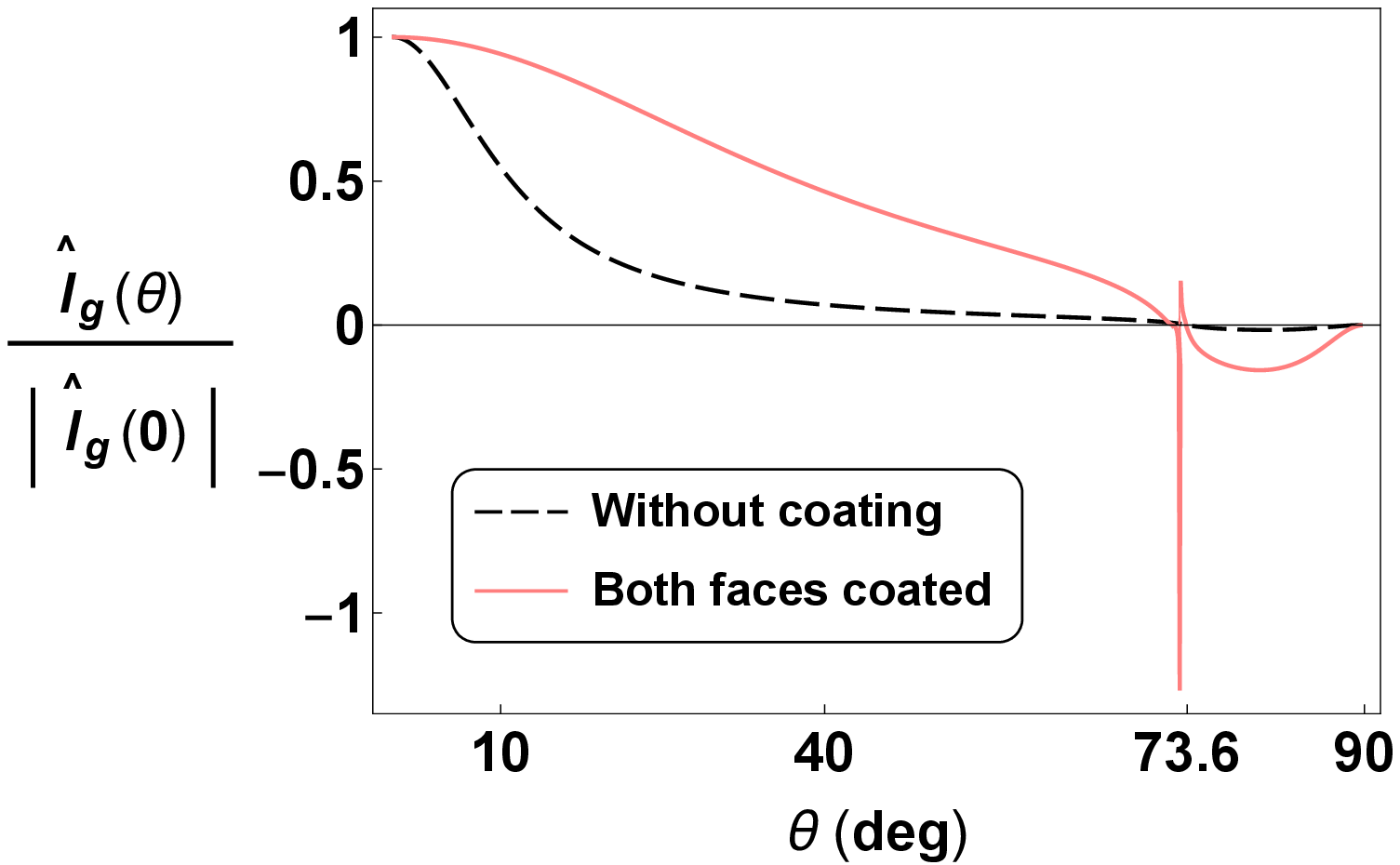}\hspace{.5cm}
         \caption{}
          \label{fig:NiceImage7}
      \end{subfigure}
      \hfill
      \begin{subfigure}{0.45\textwidth}
      \centering
       \captionsetup{justification=centering}
        \includegraphics[scale=.50]{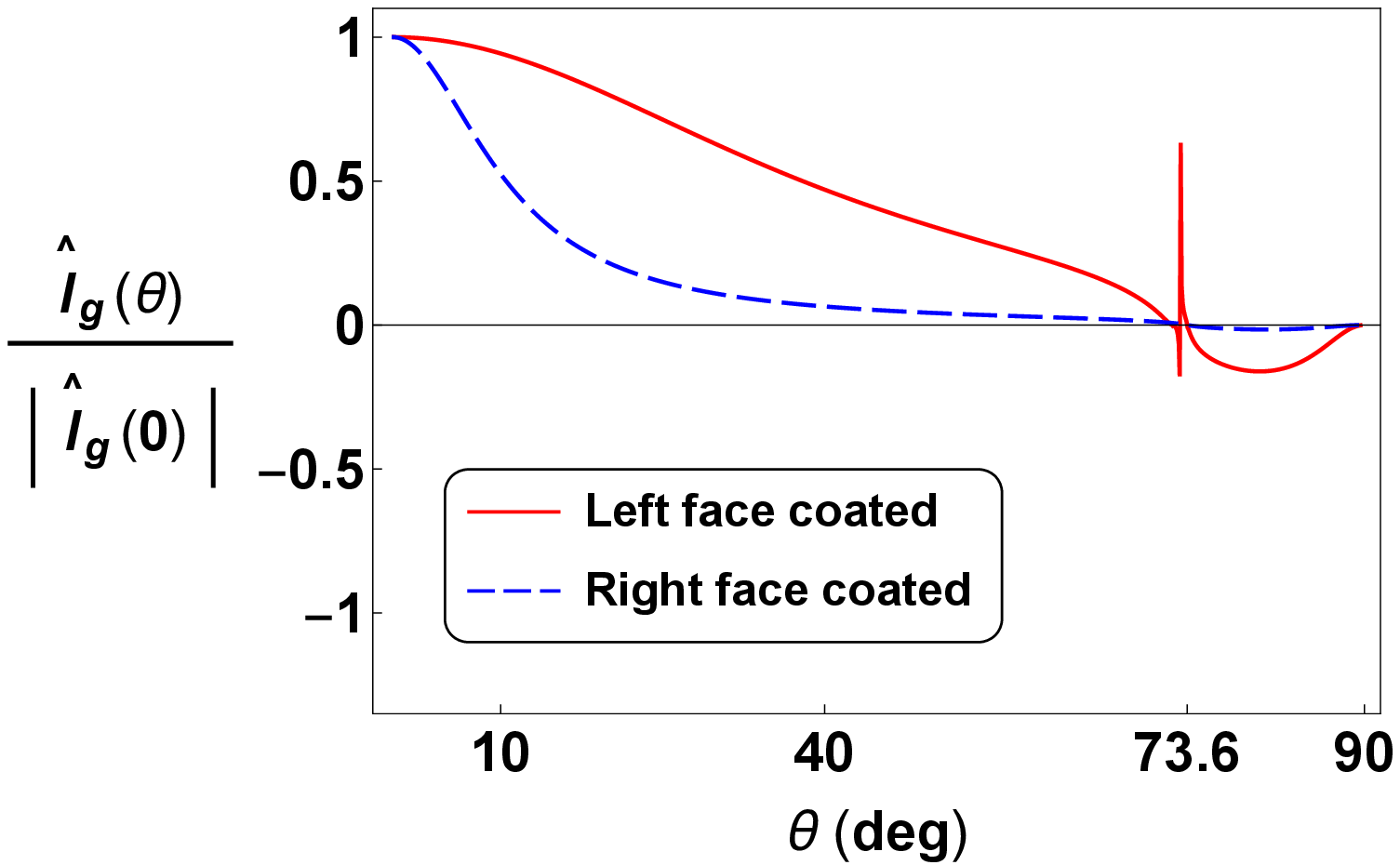}
          \caption{}
          \label{fig:NiceImage8}
      \end{subfigure}\\
           \begin{subfigure}{0.45\textwidth}
            \centering
       \captionsetup{justification=centering}
        \includegraphics[scale=.50]{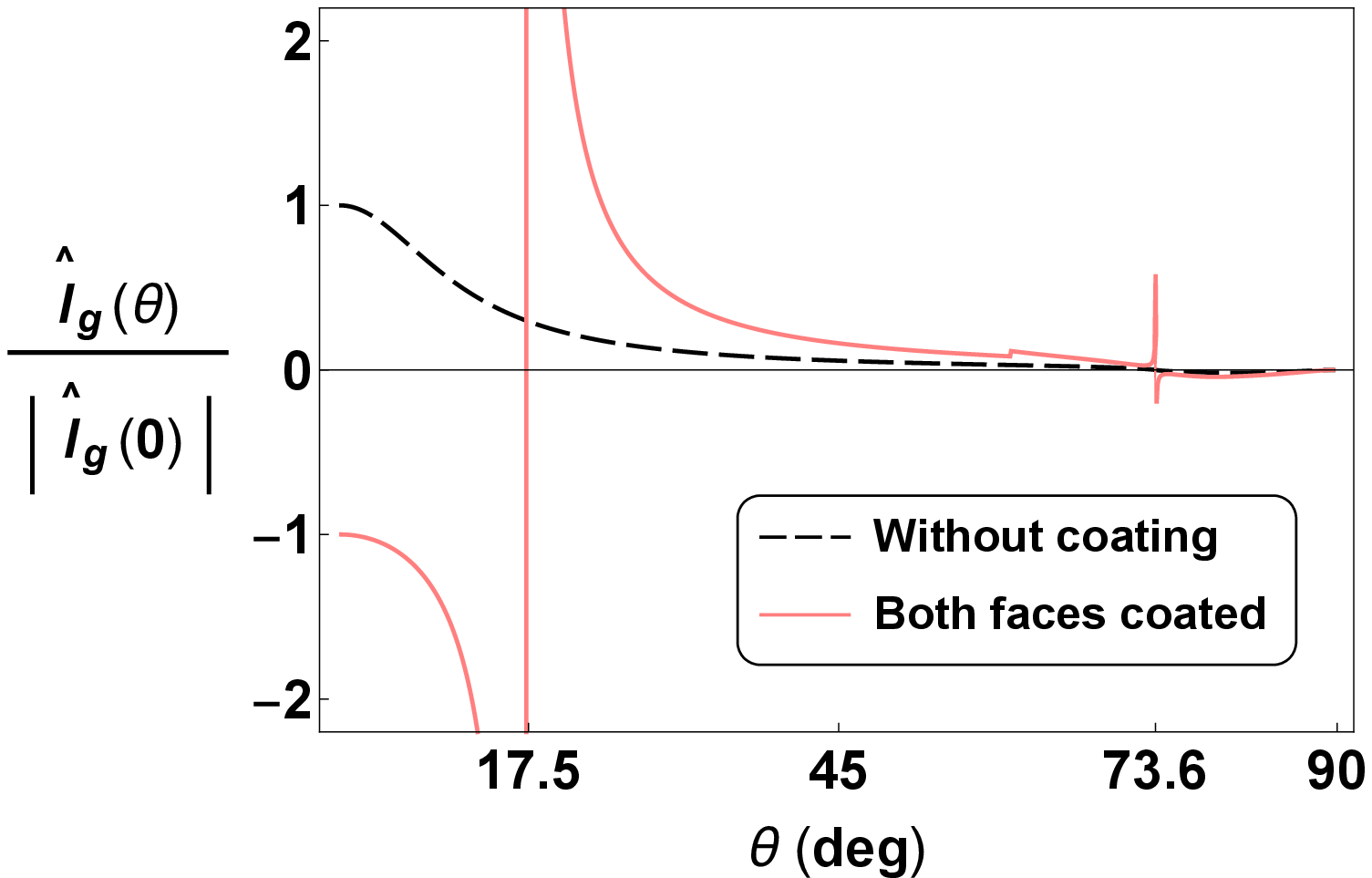}\hspace{.5cm}
          \caption{}
          \label{fig:NiceImage9}
      \end{subfigure}
      \hfill
      \begin{subfigure}{0.45\textwidth}
      \centering
       \captionsetup{justification=centering}
        \includegraphics[scale=.50]{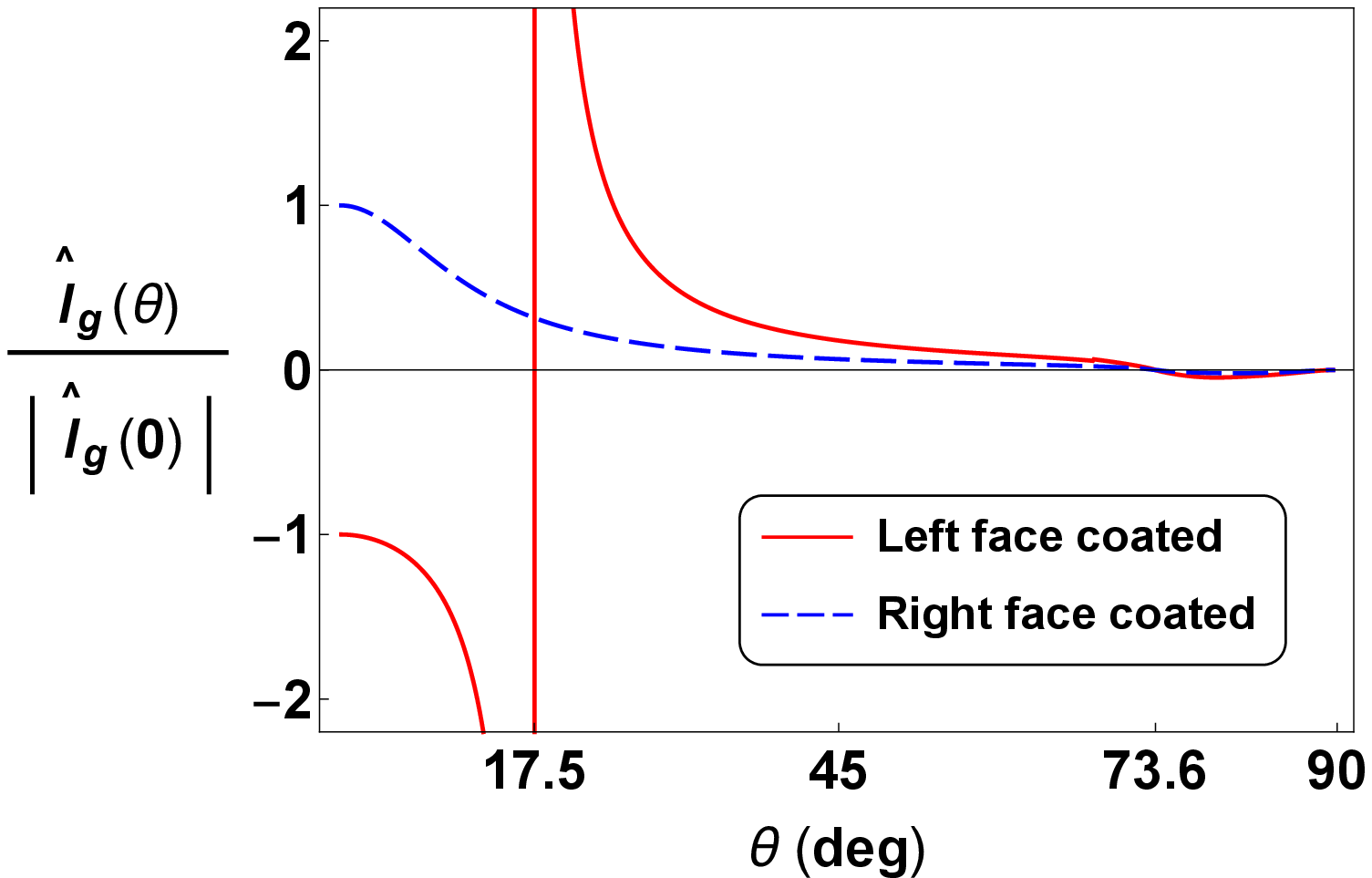}
          \caption{}
          \label{fig:NiceImage10}
      \end{subfigure}
    \caption{Graphs of $ \hat{I}_{g}(\theta)/|\hat{I}_{g}(0)|$ for lasing in the TM modes of a slab with and without Graphene (Figs.~\ref{fig:NiceImage7} and \ref{fig:NiceImage8}) or WSM (Figs.~\ref{fig:NiceImage9} and \ref{fig:NiceImage10}) coatings for $L=300\,\mu{\rm m}$, $\eta_0=3.4$, $\lambda_0=1.5\,\mu{\rm m}$, and $T_0=300^\circ K$. $\theta_\star=17.5^\circ$ marks the singularity of $\hat{I}_{g}(\theta)$ when the slab's left or both faces are coated by WSM. $\theta_b=73.6^\circ$ is the Brewster's angle. Values of $\hat{I}_{g}(0)$ are listed in Table~\ref{table1}.}
    \label{IgTM-WSM}
    \end{center}
    \end{figure*}

Fig.~\ref{IgTM-WSM} shows the graphs of the normalized intensity slope $ \hat{I}_{g} $ as a function of incidence angle $ \theta $ in the TM modes of a slab with and without Graphene or WSM coatings. According to these figures coating the right face of the slab has negligible effect on the value of $\hat{I}_{g}$. Furthermore, the presence of the coating on one or both faces of the slab does not change the fact that for emission angles $\theta$ exceeding the Brewster's angle, $ \hat{I}_{g}$ takes a negative value. This in particular forbids lasing in the TM modes of the slab for emission angles exceeding the Brewster's angle unless we can adjust $\eta, T$, and $b'$ properly. Note however that the latter will prevent lasing in the TM modes with $\theta<\theta_b$.

For a slab with WSM coating on its left or both faces, $\hat{I}_{g}(\theta)$ diverges at a critical angle $\theta_\star$. It takes negative (respectively positive) values for $\theta<\theta_\star$ and $\theta>\theta_b$ (respectively $\theta_\star<\theta<\theta_b$.) As explained in ref.~\cite{jo-2017}, the same phenomenon appears for a slab without coatings, if $\eta_0\leq 3$. However, here the singularity turns out to persist for all values of $\eta_0$. This suggests that for TM modes with emission angle $\theta=\theta_\star$ our perturbative analysis is inconclusive. For $\theta$ values substantially smaller than $\theta_\star$, $\hat{I}_{g}(\theta)$ takes a finite and negative value. This is again an indiction that unless we properly tune $\eta$ and $b'$, we can not induce laser emission in the corresponding TM modes of the slab from its right-hand face.

Next, we examine the behavior of the ratios of the intensity slops $\cX_\eta:=\hat I_\eta/\hat I_g$, $\cX_T:=\hat I_T/\hat I_g$, and $\cX_{b'}:=\hat I_{b'}/I_g$ that enter the expression (\ref{intensity-relation-n}) for the output intensity of our slab laser.

\begin{figure*}
    \begin{center}
    \begin{subfigure}{0.45\textwidth}
      \centering
       \captionsetup{justification=centering}
        \includegraphics[scale=.43]{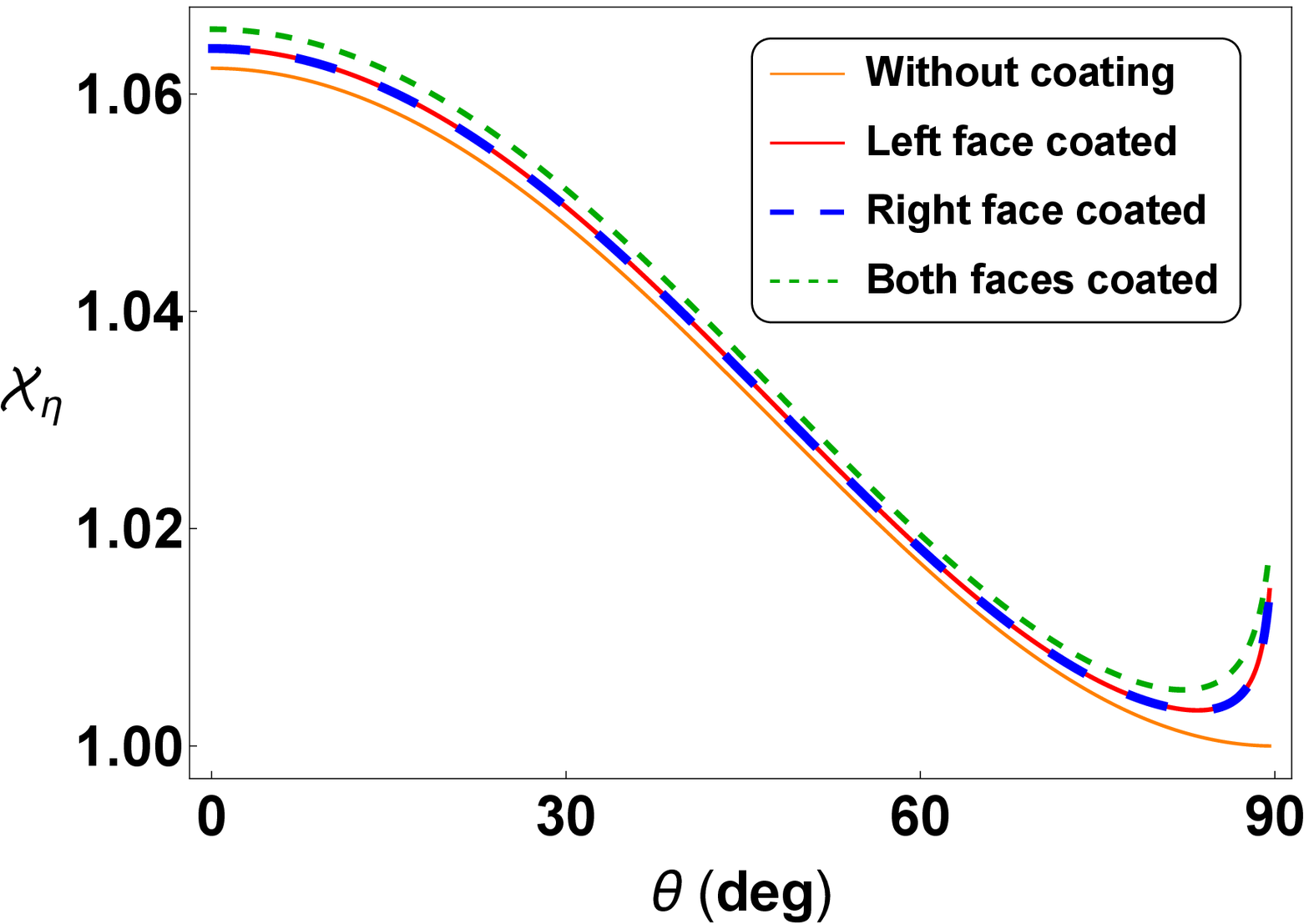}\hspace{0.5cm}
         \caption{}
          \label{fig:NiceImage11}
      \end{subfigure}
      \hfill
      \begin{subfigure}{0.45\textwidth}
      \centering
       \captionsetup{justification=centering}
        \includegraphics[scale=.43]{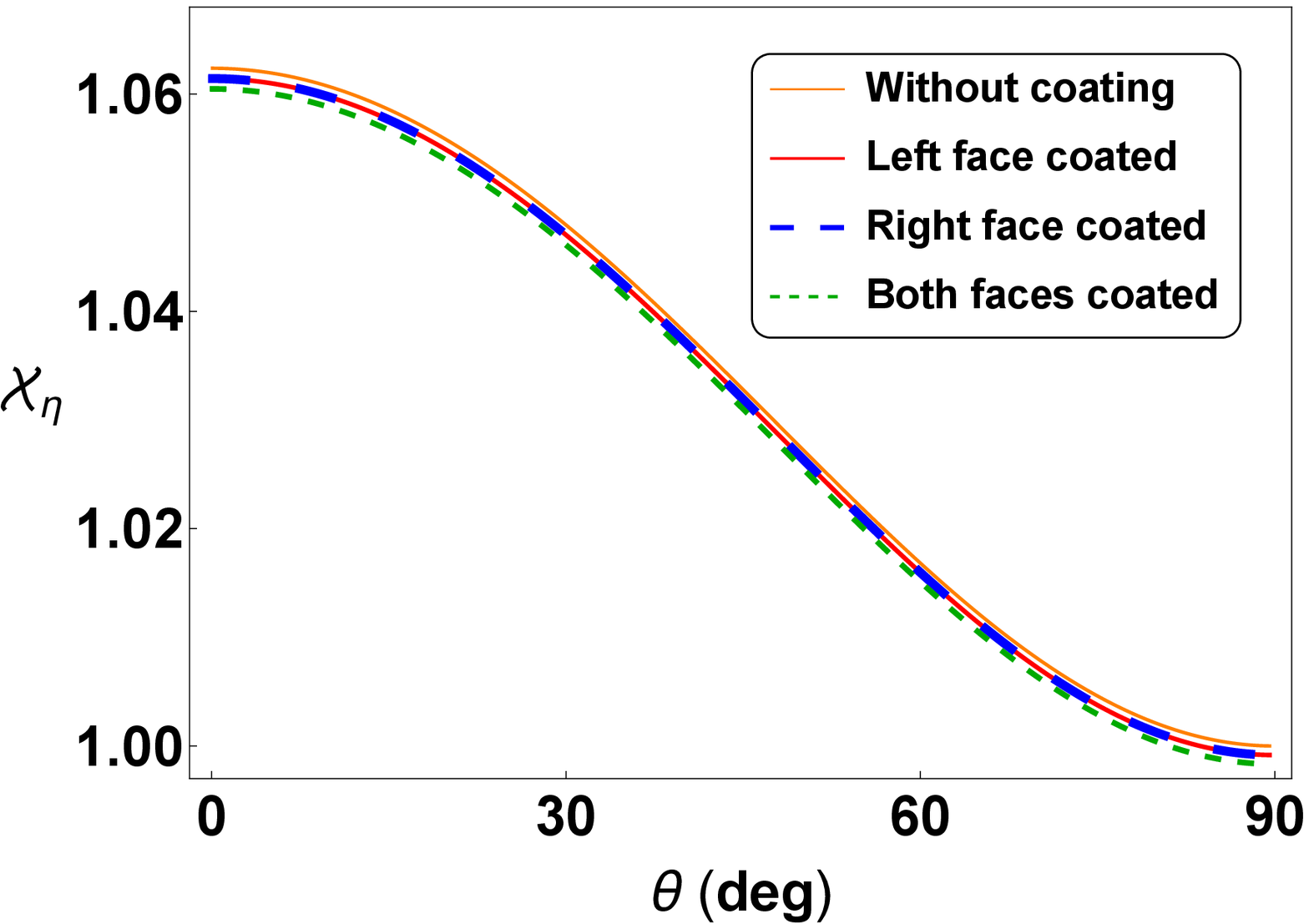}
          \caption{}
          \label{fig:NiceImage12}
      \end{subfigure}\\[6pt]
           \begin{subfigure}{0.45\textwidth}
            \centering
       \captionsetup{justification=centering}
        \includegraphics[scale=.42]{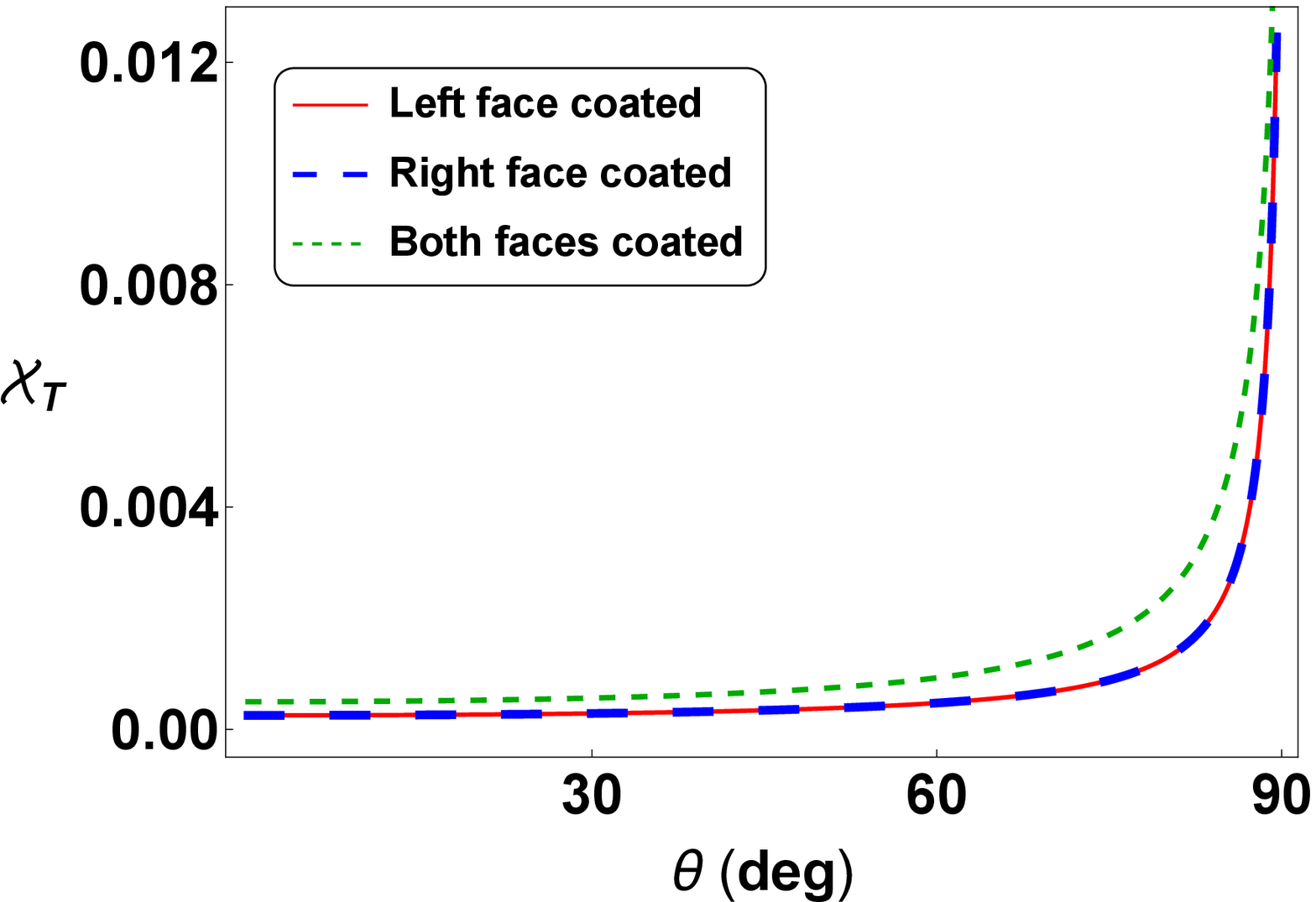}\hspace{0.5cm}
          \caption{}
          \label{fig:NiceImage13}
      \end{subfigure}
      \hfill
      \begin{subfigure}{0.45\textwidth}
      \centering
       \captionsetup{justification=centering}
        \includegraphics[scale=.45]{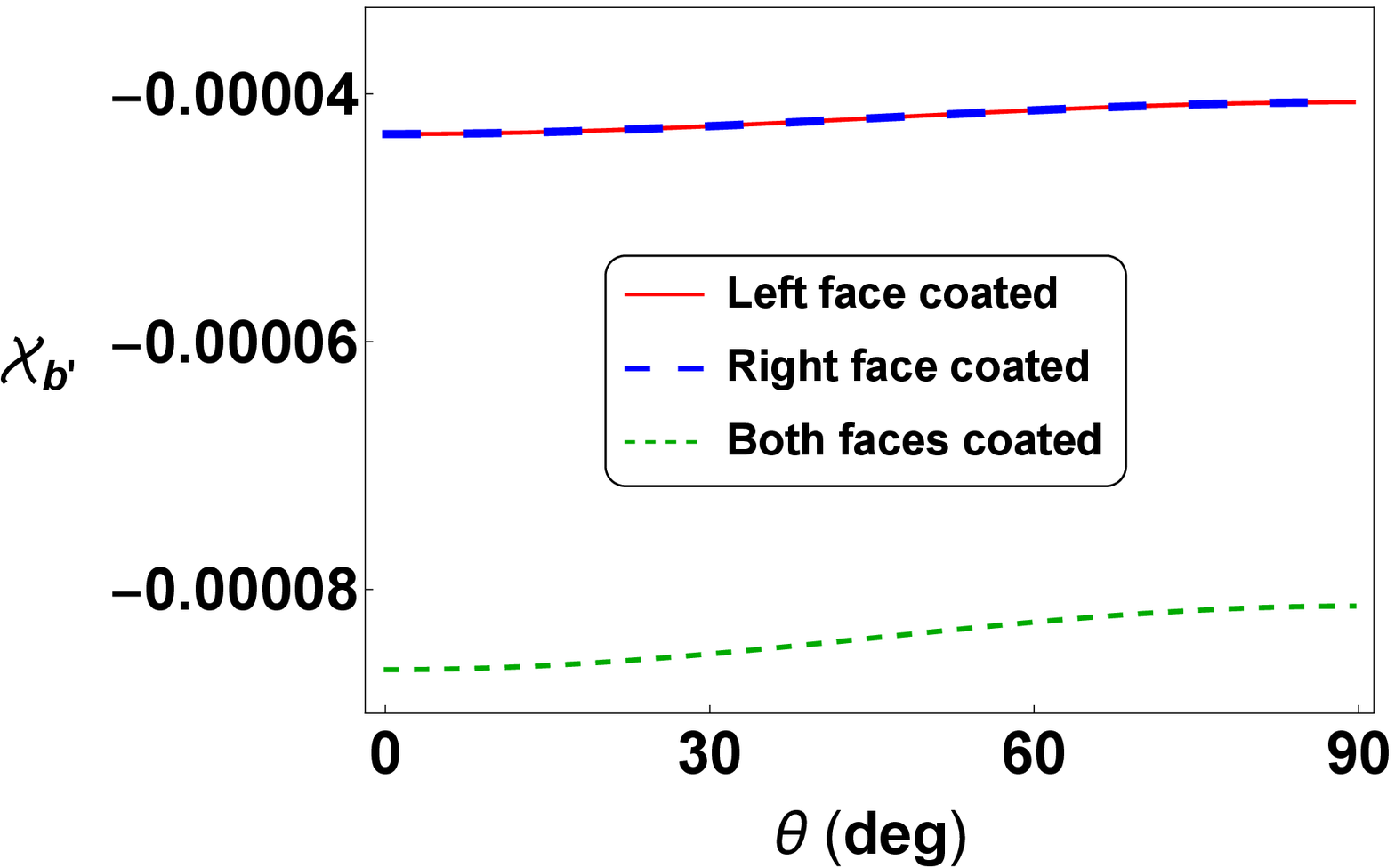}
          \caption{}
          \label{fig:NiceImage14}
      \end{subfigure}
    \caption{Graphs of $\cX_\eta$, $\cX_T$, and $\cX_{b'}$ as functions of $\theta$ for lasing in the TE modes of a slab with and without Graphene (Figs.~\ref{fig:NiceImage11} and \ref{fig:NiceImage13}) or WSM (Figs.~\ref{fig:NiceImage12} and \ref{fig:NiceImage14}) coatings for $L=300\,\mu{\rm m}$, $\eta_0=3.4$, $\lambda_0=1.5\,\mu{\rm m}$, $ b'_0=0.0005~{\angstrom} $, and $T_0=300^\circ K$.}
    \label{X-TE}
    \end{center}
    \end{figure*}

Fig.~\ref{X-TE} shows the graphs of $\cX_\eta$, $\cX_T$, and $\cX_{b'}$ as functions of $\theta$ for lasing in the TE modes of a slab with and without Graphene or WSM coatings. The presence of the coatings has a negligible effect on the value of  $\cX_\eta$. This is a clear evidence that the idea of boosting the intensity of the laser by tuning the parameters of the system requires adjusting the parameters entering the conductivity of the coating layers, namely $T$ for Graphene and $b'$ for WSM, rather than the real part of the refractive index of the slab, $\eta$. Notice also that because $\cX_{b'}$ takes extremely small values, this scheme is more effective for a slab with Graphene coatings on its faces and TE modes with larger emission angle.

\begin{figure*}
    \begin{center}
    \begin{subfigure}{0.45\textwidth}
      \centering
       \captionsetup{justification=centering}
        \includegraphics[scale=.43]{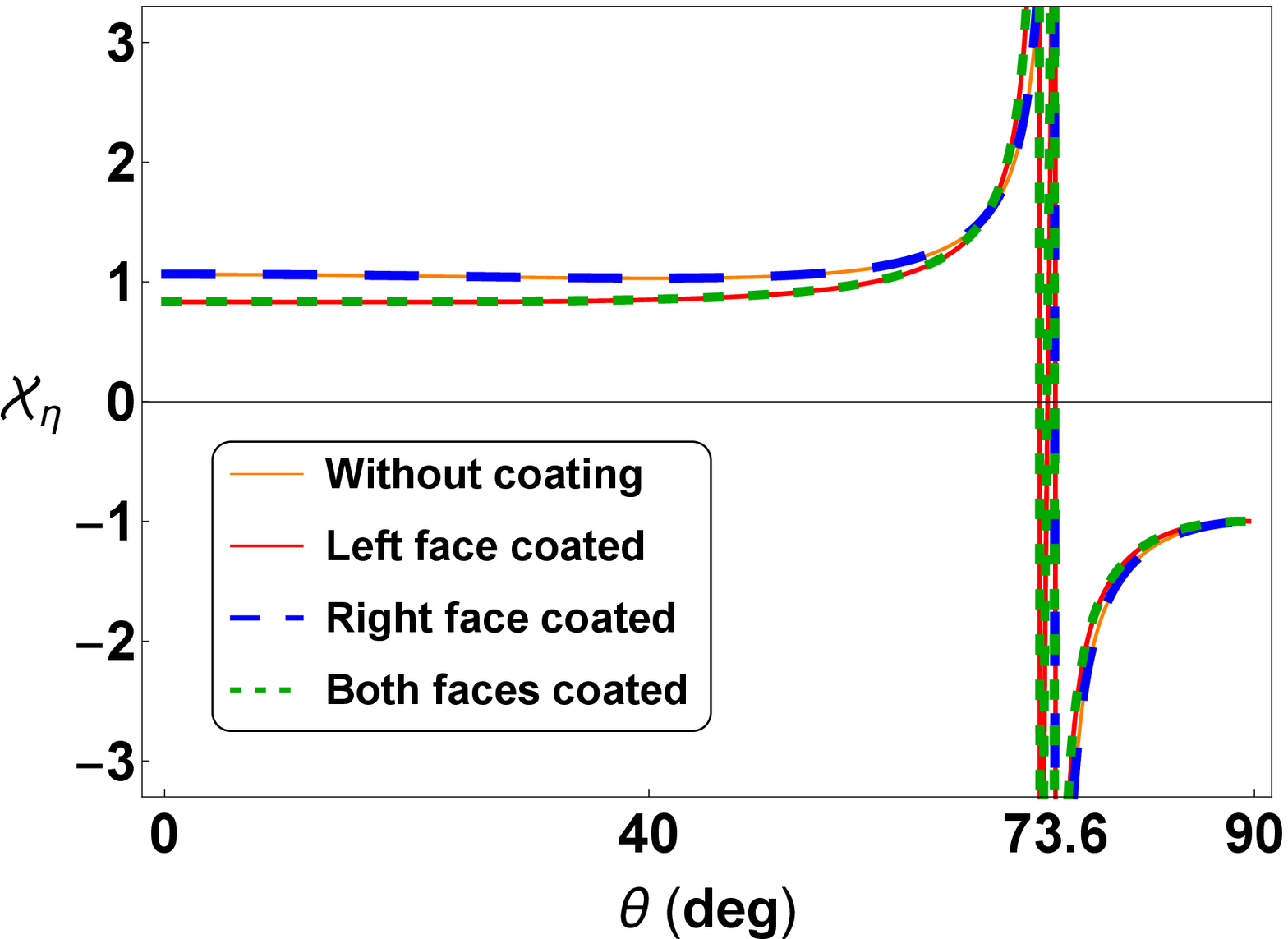}\hspace{0.5cm}
         \caption{}
          \label{fig:NiceImage15}
      \end{subfigure}
      \hfill
      \begin{subfigure}{0.45\textwidth}
      \centering
       \captionsetup{justification=centering}
        \includegraphics[scale=.43]{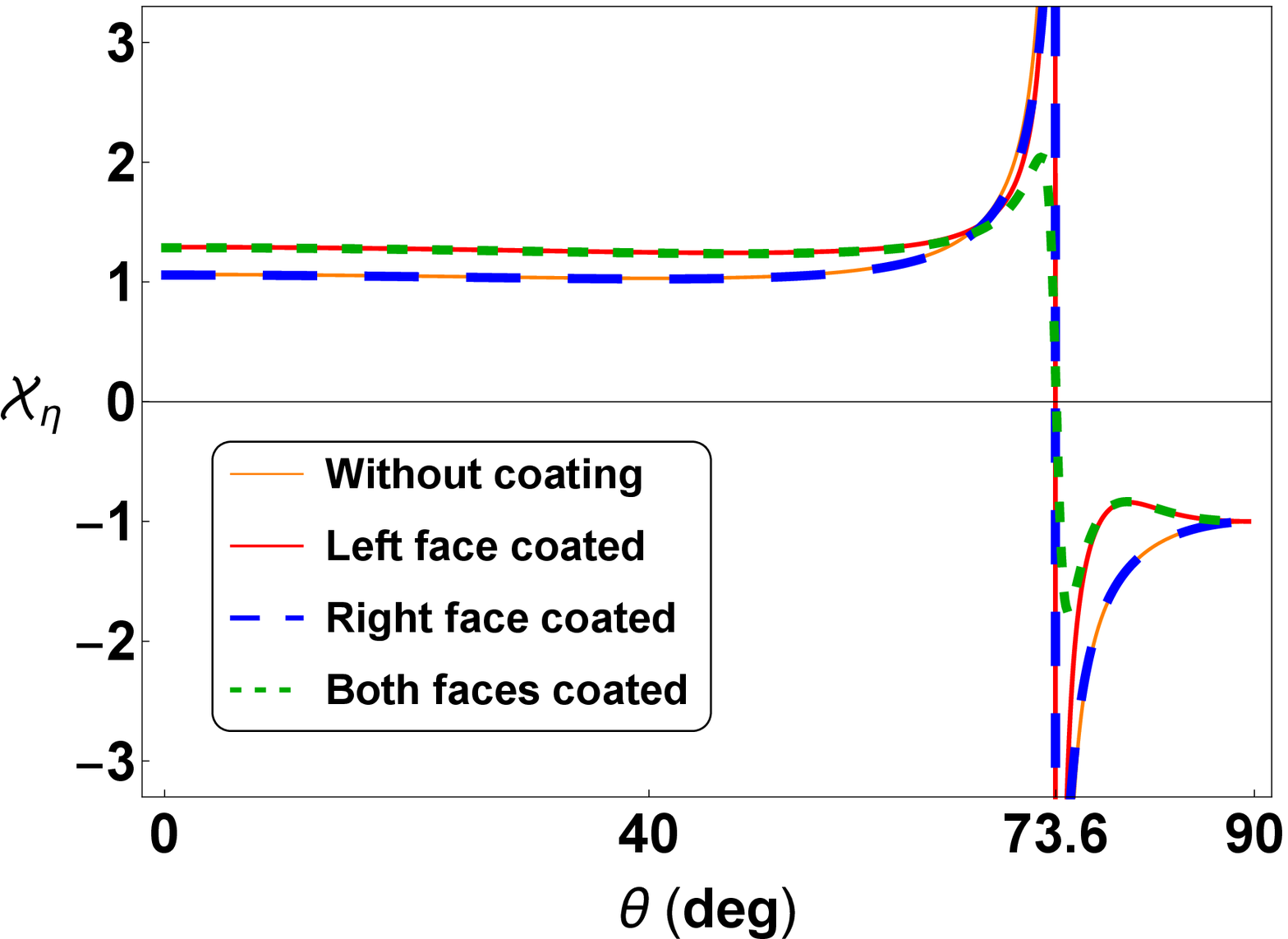}
          \caption{}
          \label{fig:NiceImage16}
      \end{subfigure}\\[6pt]
           \begin{subfigure}{0.45\textwidth}
            \centering
       \captionsetup{justification=centering}
        \includegraphics[scale=.42]{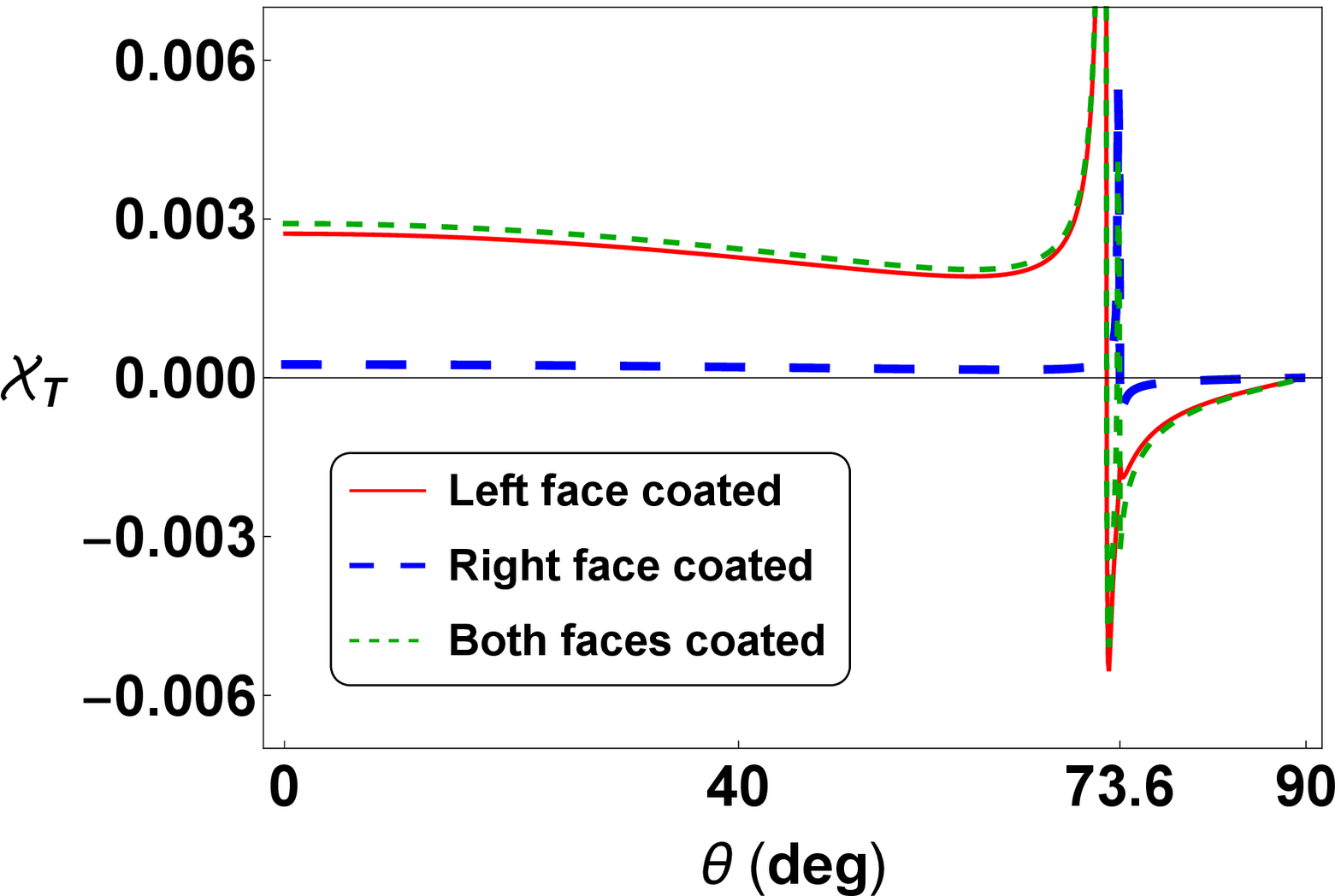}\hspace{0.5cm}
          \caption{}
          \label{fig:NiceImage17}
      \end{subfigure}
      \hfill
      \begin{subfigure}{0.45\textwidth}
      \centering
       \captionsetup{justification=centering}
        \includegraphics[scale=.43]{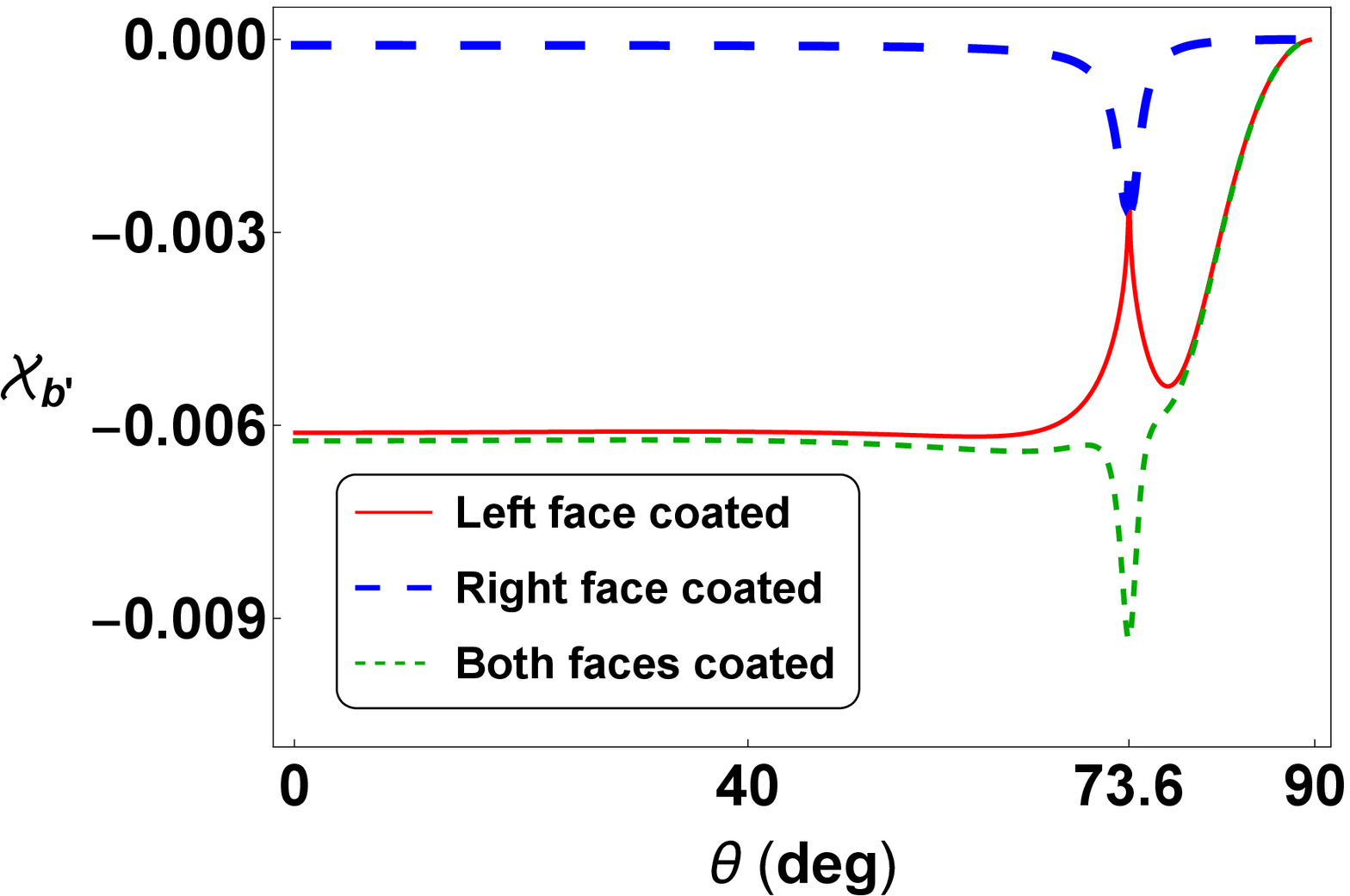}
          \caption{}
          \label{fig:NiceImage18}
      \end{subfigure}
    \caption{Graphs of $\cX_\eta$, $\cX_T$, and $\cX_{b'}$ as functions of $\theta$ for lasing in the TM modes of a slab with and without Graphene (Figs.~\ref{fig:NiceImage15} and \ref{fig:NiceImage17}) or WSM (Figs.~\ref{fig:NiceImage16} and \ref{fig:NiceImage18}) coatings for $L=300\,\mu{\rm m}$, $\eta_0=3.4$, $\lambda_0=1.5\,\mu{\rm m}$, $ b'_0=0.0005~{\angstrom}$, and $T_0=300^\circ K$.}
    \label{X-TM}
    \end{center}
    \end{figure*}

Fig.~\ref{X-TM} gives the analog of the graphs given in
Fig.~\ref{X-TE} for the TM modes of the slab in the absence and
presence of the coatings. According to these graphs, coating the
faces of the slab does not have a sizable effect on $\cX_\eta$
except near the Brewster's angle $\theta_b$ where it has a high peak
and a low minimum. In view of the fact that $g_0$ becomes
unrealistically large in the vicinity of $\theta_b$ and $\hat I_g$
takes a negative value for $\theta>\theta_b$, we conclude that
coating the faces of the slab does not remove the obstruction on the
emission of TM waves with $\theta\geq\theta_b$. Furthermore, for a
slab with WSM coatings, $\cX_\eta$ and $\cX_{b'}$ are smooth
functions at the singular angle $\theta_\star$ while $\hat I_g$
takes a negative value for angles $\theta<\theta_\star$ and
$\cX_\eta$ is nealy 1 for these angles. These observations suggest
that lasing in TM modes with $\theta\leq\theta_\star$ is also
forbidden for this slab. Another distinctive feature of the TM modes
is that the Graphene and WSM coatings on the left or both faces of
the slab lead to values of $|\cX_T|$ and $|\cX_{b'}|$ that are of
the same order of magnitude. Hence in principle one expects to be
able to boost the output intensity of the slab laser with either
type of coatings.

   For the slab with WSM coatings, the dependence of the laser output intensity on the distance between Weyl nodes $b'$ suggests using lasing characteristics of the slab laser for the purpose of identifying the value of $b'$. This may provide the basis for a method of determining $b'$ which is currently a major experimental challenge. Let us also mention that in case of coatings with graphene sheets, we can also consider the chemical potential $\mu$ as a tunable parameter. It turns out, however, that changing $\mu$ has a negligible effect on the output intensity. Therefore, we do not include the related graphical data here.

\section{Summary and conclusion}

The fact that laser emission corresponds to the scattering solutions of the relevant wave equation with the purely outgoing boundary conditions \cite{tureci-2006} and the observation that these solutions mark the emergence of spectral singularities \cite{prl-2009} underline the importance of the latter in mathematical descriptions of lasers. An important outcome of recognizing this connection is a purely mathematical prescription for deriving the laser threshold condition for lasers with a variety of  geometries \cite{pra-2011a,pla-2011,prsa-2012,pra-2013b,pra-2013d,pra-2015}. This connection has provided the motivation for a nonlinear generalization of the concept of spectral singularity \cite{prl-2013}, led to a simple mathematical derivation of the linear relationship between the laser output intensity and the gain coefficient \cite{pra-2013,jo-2017}, and offered a method for computing the intensity slope appearing in this relation. In the present paper we have used this method to explore the consequences of coating one or both faces of a mirrorless slab laser by a thin 2D material with scalar conductivity. We have obtained a generalization of the well-known linear relation between the laser output intensity and gain coefficient that takes into account the effect of the coating. In particular, the physical parameters determining the conductivity of the 2D material enter the expression for the output intensity. Therefore, by tuning these parameters we can adjust the intensity of the emitted light.

We have conducted a comprehensive study of the behavior of the intensity slopes associated with all the relevant parameters for a slab with Graphene or Weyl semimetal coatings. These in turn revealed interesting phenomena such as a splitting of the Brewster's angle associated with lasing in the TM modes of the slab and the presence of a singular emission angle $\theta_\star$ that arises in the study of the lasing in the TM modes of the slab with Weyl semimetal coatings. Although a careful description of this singularity and its physical consequences is beyond the capabilities of our perturbative analysis, our results suggest that Weyl semimetal coatings obstruct lasing in the TM modes of the slab with emission angles $\theta\leq\theta_\star$. For TM modes with $\theta$ slightly larger than $\theta_\star$, Weyl semimetal coatings cause a notable increase in the intensity slopes. This shows that for these modes small changes in the conductivity parameters of the coating can yield considerable changes in laser output intensity.

As a final comment we wish to point out that both Graphene~\cite{wright2009} and WSM~\cite{ooi-2019} exhibit strong nonlinear effects in terahertz to infrared frequencies. In the present paper, we consider lasing in the visible spectrum where we can safely ignore these nonlinearities. One can in principle study the consequences of including these nonlinearities in the calculation of linear and nonlinear spectral singularities and their effect on the threshold gain and laser output intensity in the terahertz regime.

\subsection*{Acknowledgements} This work has been supported by  the Scientific and Technological Research Council of Turkey (T\"UB\.{I}TAK) in the framework of the project no: 114F357, and by the Turkish Academy of Sciences (T\"UBA).

\ed